\documentclass[preprint,12pt]{elsarticle}

\usepackage{amssymb}

\usepackage{multirow}
\usepackage{caption}
\usepackage{subcaption}
\usepackage{xcolor}
\usepackage{url}

\usepackage{breakurl}
\usepackage[breaklinks]{hyperref}
\usepackage{color}
\definecolor{cyan}{rgb}{0.2,0.6,1}
\definecolor{red}{rgb}{1,0,0}
\definecolor{blue}{rgb}{0,0,1}
\definecolor{green}{rgb}{0,0.7,0}
\usepackage{ulem}
\journal{}
\usepackage{nomencl}
\makenomenclature
\begin{document}

\begin{frontmatter}

\title{The indoor agriculture industry: a promising player in demand response services}

\author[inst1]{Javier Penuela}
\author[inst2]{C\'ecile Ben}
\author[inst2]{Stepan Boldyrev}
\author[inst2]{Laurent Gentzbittel}
\author[inst1]{Henni Ouerdane}

\affiliation[inst1]{department={Center for Digital Engineering,}, organization={Skolkovo Institute of Science and Technology},
            addressline={30 Bolshoy Boulevard}, 
            city={Moscow},
            postcode={121205},
            country={Russia}}

\affiliation[inst2]{department={Project Center for Agro Technologies,}, organization={Skolkovo Institute of Science and Technology},
            addressline={30 Bolshoy Boulevard}, 
            city={Moscow},
            postcode={121205},
            country={Russia}}

\begin{abstract}
Demand response (DR) programs currently cover about 2\% of the average annual global demand, which is far from contributing to the International Energy Agency's ``Net Zero by 2050'' roadmap's 20\% target. While aggregation of many small flexible loads such as individual households can help reaching this target, increasing the participation of industries that are major electricity consumers is certainly a way forward. The indoor agriculture sector currently experiences a significant growth to partake in the sustainable production of high-quality food world-wide. As energy-related costs, up to 40\% of the total expenses, may preclude full maturity of this industry, DR participation can result in a win-win situation. Indeed, the agriculture system must transform and become a sustainable source of food for an increasing number of people worldwide under the constraints of preservation of soils and water, carbon footprint, and energy efficiency. We considered the case of the Russian Federation where indoor farming is burgeoning and already represents a load of several thousand megawatts. To show the viability of the indoor farming industry participation in implicit and explicit DR programs, we built a physical model of a vertical farm inside a phytotron with complete control of environmental parameters including ambient temperature, relative humidity, CO$_2$ concentration, and photosynthetic photon flux density. This phytotron was used as a model greenhouse. We grew different varieties of leafy plants under simulated DR conditions and control conditions on the same setup. Our results show that the indoor farming dedicated to greens can participate in DR without adversely affecting plant production and that this presents an economic advantage.
\end{abstract}




\end{frontmatter}



\nomenclature{DR}{Demand response}
\nomenclature{$A$}{Total illuminated surface area [m$^2$]}
\nomenclature{PAR}{Photosynthetic active radiation}
\nomenclature{DLI}{Daily light integral}
\nomenclature{$C_i$}{Russian spot market electricity price on a given day at a given hour $i$ [rub$\cdot$MWh$^{-1}$]}
\nomenclature{$\eta_i$}{Hourly averaged photosynthetic photon flux density at a given hour $i$ [mol$\cdot $m$^{-2}\cdot$h$^{-1}$]}
\nomenclature{$R_i$}{Average hourly photosynthetic active radiation at a given hour $i$ [W$\cdot$m$^{-2}$]}
\nomenclature{$k_{\rm ph}$}{= $6.12\times 10^3$  mol$\cdot$ MWh$^{-1}$ - measured photon efficacy of our LED fixture.}
\nomenclature{$k_{\rm PAR}$}{ = 2.02 - photosynthetic photon flux density conversion coefficient from W$\cdot$m$^{-2}$ to mol$\cdot\mu$m$^{-2}\cdot$s$^{-1}$ from solar radiation}
\nomenclature{$K_{\rm trans}$}{ = 0.8 - light transmission coefficient of the greenhouse walls.}
\nomenclature{$\rho_i$}{Calculated photsynthetic photon flux density inside the greenhouse due to solar radiation at a given hour $i$ [W $\cdot$ m$^{-2}$]}
\nomenclature{PPFD}{Photosynthetic photon flux density [W$\cdot$m$^{-2}$]}
\nomenclature{PWM}{pulse-width modulation}
\nomenclature{$R_{\rm opt}$}{Optimal daily light integral [mol$\cdot $m$^{-2}\cdot $day$^{-1}$]}
\nomenclature{$h$}{Daily photoperiod [h]}
\nomenclature{HPS}{High pressure sodium lamp}
\nomenclature{$P_{\rm cons}$}{Measured averaged power consumption of the LEDs [W]}

\printnomenclature

\section{Introduction}
\label{intro}
The stability of power grids is key to guarantee the reliable and safe distribution of electricity to consumers. However, risks of contingency are many as they can arise because of a sudden demand increase, the intermittent character of renewable energy sources, electric faults, or the uncertainty of the power market. To meet high demand, peaking power plants can provide the needed surplus of electricity within a few minutes of start-up but at an increased price because of the costs incurred by the operation of such generators in addition to the base load plant operation. Indeed, peaking power generation is less efficient, causes more pollution, and can lead to grid congestion and hence increased dissipation of electrical energy into waste heat. Besides, a too-low consumption (in case of significant renewable penetration with low or insufficient storage capacity) also results in energy loss and grid imbalance  \cite{albadi2008summary,siano2014demand}. Demand response (DR) programs form one approach to mitigate risks of instability by maintaining the balance between electricity production on the generation side and the consumption on the load side \cite{albadi2008summary,siano2014demand}.

\begin{figure}[h]
		\centering
		\includegraphics[width=0.9\linewidth]{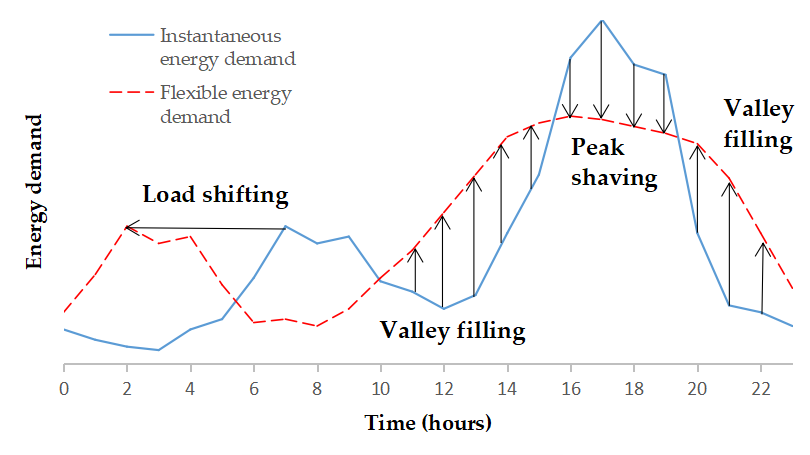}
		\caption{Illustration of the different possible strategies that can be accomplished using DR.}
		\label{fig:demand-side-management-energy}
\end{figure}

The basic principle of DR is the provision by the grid operator of incentives to customers to reduce their demand of electricity during peak time \cite{christensen2020agent}. This usually works in two different ways as illustrated in Fig. \ref{fig:demand-side-management-energy}. One way is implicit DR, which affects the demand by changing the hourly electricity prices, thus resulting in load shifting: the customers may decide to consume electricity when it is cheaper. The other is explicit DR, in which case the grid operator sends a signal to the consumer to either disconnect a beforehand agreed load (peak shaving) or increase the consumption (valley filling). The accumulated cost of the rewards and price reductions is several times lower than that of building new infrastructures \cite{albadi2008summary,VygonDR2018} thus limiting the ecological impact of the grid.
Note that the development of modern DR programs entails more grid flexibility to allow full integration of new participants, additional energy sources, two-way energy flow, and minimal reliance on peaking generation units and storage facilities that require additional transmission and distribution infrastructures \cite{albadi2008summary}. This, however, necessitates sensors and information technologies distributed across the power system to make the grid smart, efficient and reliable \cite{christensen2020agent,christensen2019agent}. Demand-side flexibility is one of the already available and relatively cheap smart-grid solutions, which in addition to DR, includes coupling different energy sectors such as power-to-heat, power-to-hydrogen, electric-vehicle charging, and smart appliances \cite{irena2019demand}. To target new players who are ready to operate in a more flexible fashion, DR models must demonstrate that participation has no adverse effect on product quality and business operation.

Turning to agriculture, practices have evolved much since its inception, tending to productivity increase and better use of resources fostered by innovation and available technologies. However, agriculture has gradually become a source of ecological problems including soil erosion, water depletion, land subsidence, and destruction of cropland~\cite{periard2012irrigation, branca2013food,Environment2021Stein, BLOM2022134443}. Whether extensive, intensive or organic, each type of agriculture impacts in a different fashion on the environment \cite{periard2012irrigation,branca2013food, BLOM2022134443}. Industrial greenhouses and vertical farms have been proposed among the solutions to the ecological challenges posed by intensive agriculture. Their main characteristics are the usage of a closed volume for plant growth and the control of all the indoor environmental factors \cite{Environment2021Stein,vanDelden2021,alshrouf2017hydroponics,buildings8020024}. Indoor farming benefits therefore include a lower ecological impact, increased food safety, promotion of the local economy, stabilization and reduction in the price of vegetables during off-season time \cite{vanDelden2021,buildings8020024}. The best results are seen in hydroponic technologies  \cite{Environment2021Stein,alshrouf2017hydroponics}, which make plant growth possible in soil-less media such as, e.g., sand, water, air. These technologies are mainly used for vertical agriculture with rows of plants placed above the ground level, optimizing the use of the whole volume in the controlled environment \cite{Environment2021Stein,vanDelden2021,Futurefood2017Kurt}. However, while the indoor agriculture industry has the lowest impact on resources and environment contamination, its sustainability can be questioned given the high carbon emissions \cite{BLOM2022134443}. For instance, the production of one kilogram of butter head lettuce in vertical farming can generate up to 2.5 times more CO$_{2}$-equivalent emissions than in industrial greenhouses, and 17 times more than in traditional open field agriculture  \cite{BLOM2022134443}.
	
Control of environmental factors such as sun light, humidity, and carbon dioxide concentration, is a key feature of the most economically-efficient models of indoor farming \cite{Environment2021Stein}. Yet, these factors are the cause of a significant dependence on the electric grid: in high latitude locations, about 70\% of the power consumption in indoor farming is for light, and about 28\% for humidity and temperature control \cite{BLOM2022134443,Verticalfarm2,sorensen2016dynagrow,ma2018energy}. Overall, energy expenses can reach 40\% of the total production cost, making these quite significant, if not excessive for the long-term growth and full maturation of the indoor agriculture industry \cite{Environment2021Stein,Wheat}. On the other hand, DR programs suffer from slow adoption rates by customers: they currently cover about 2\% of the average annual global demand, and their implementation is reported by the International Energy Agency as insufficient to reach the ``Net Zero by 2050''. Concretely, this implies a commitment by customers in economically developed countries to reach a rate of 25\% of the average annual demand, and 15\% in developing countries \cite{iea_2021,iea_2020}. The present work is a contribution to bridging this gap, focusing on the integration of DR in the currently fast-growing indoor farming industry.

In this article, we ask if the indoor farming industry, whose share currently increases in the agriculture sector of the economy, can become an important participant in DR programs. We assess and discuss the mutual benefits of participating in DR programs so that this energy management strategy may become a new standard for indoor farming \cite{christensen2019agent,Environment2021Stein,sorensen2016dynagrow,neoanalytics_2020}. We focus on the electric load related to complementary lighting excluding the thermal load as accounting for it presents several challenges \cite{LUCHNIKOV2021116419, SLEPTSOV2021216}, which necessitate a separate work. To study the flexibility of the indoor farming load and establish its optimal participation in DR programs, we performed a series of experiments, focusing on the growth of leafy plants. The national context for our simulations is that of the Russian Federation, a country where the agriculture sector is growing at a fast pace while the electrical grid is being modernized in the frame of the digitization of the industry \cite{Alekseev_Lobova_Bogoviz_Ragulina_2019}. Using our experimental results and performing a techno-economic estimation lead us to conclude that integration into a DR program can produce a significant positive economic impact on both the indoor farming industry and the electrical grid up to several hundreds of millions of rubles per year (circa 10 million USD). Considering the current energy market policies and available data in the Russian Federation, we also find that participation in DR does not result in a carbon emissions reduction, which might appear counter-intuitive at first glance.

The article is organized as follows. Section 2 is devoted to the presentation of the experimental design and setup of our physical model of a vertical farm in a phytotron. In section 3, we process and perform a statistical analysis of the data (fresh biomass, dry biomass, and quantity of true leaves) we obtained under different experimental conditions simulating a real-life scenario of DR. In section 4, we perform a techno-economic assessment of the participation of the indoor farming industry in DR programs in Russia including a carbon footprint analysis. The article ends with concluding remarks and highlights of the main results and their implications for the vertical farming industry.


\section{Materials and methods}
\label{exper}
We performed three experiments in the Project Center for Agro Technologies at Skoltech. Experiments I and II (EI and EII) were performed in a small-scale vertical farm we assembled and embedded in a phytotron, which provides a fully-controlled indoor environment; experiment III (EIII) also performed in a phytotron was mimicking an industrial greenhouse environment for comparison with the vertical farm. In EIII, we considered genetic variability to test if our conclusions can be generalized to a larger set of green leafy plants. 
Hence the body of data we obtained using the two types of setup is suitable to discuss the integration of the two main indoor farming technologies in DR programs.

\subsection{Setup main characteristics}

\begin{figure}[h]
\begin{subfigure}{.45\textwidth}
  \centering
	\includegraphics[width=\linewidth]{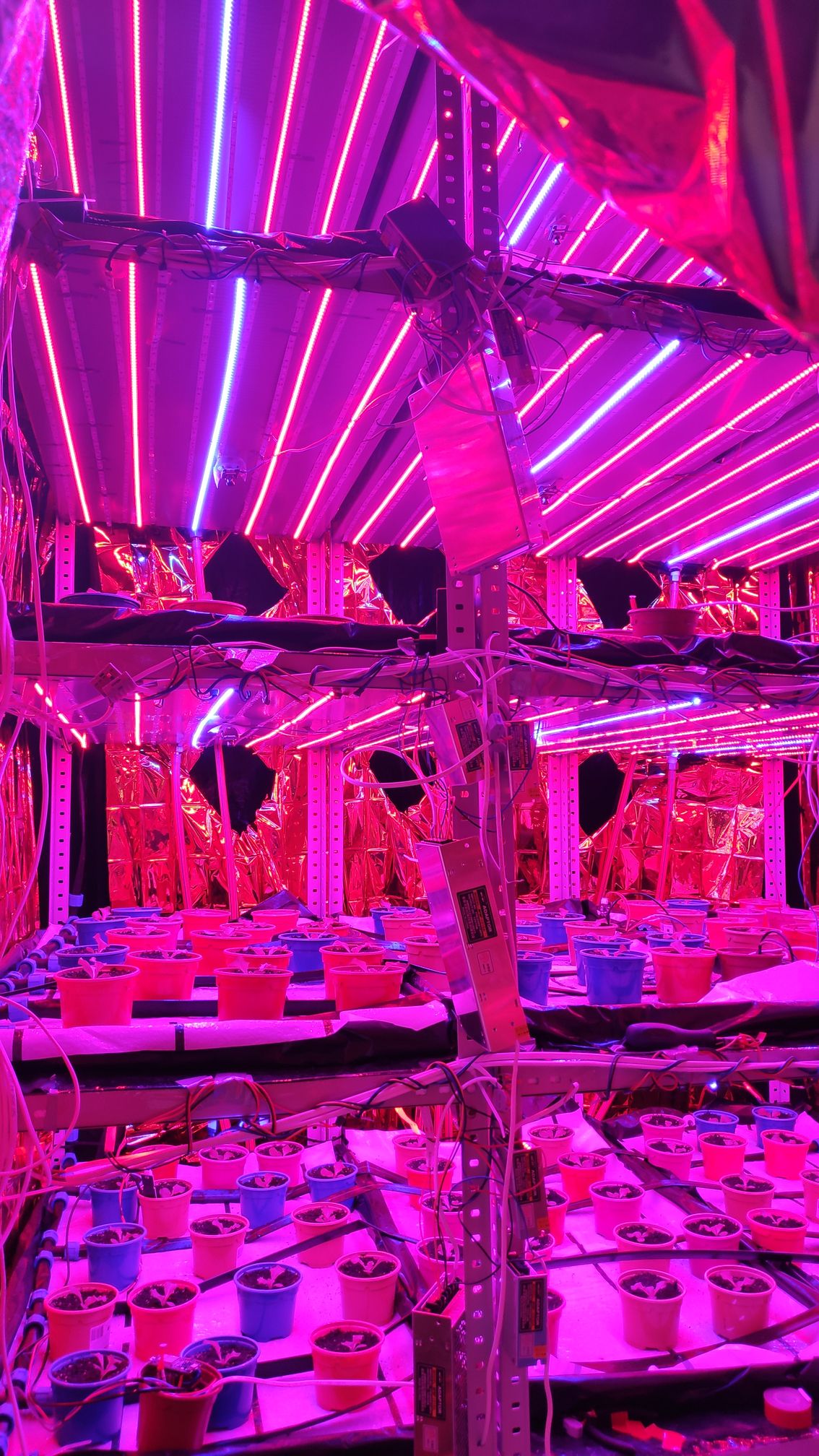}
  \caption{}
  \label{fig:setupin}
\end{subfigure}
\centering
\begin{subfigure}{.45\textwidth}
  \centering
\includegraphics[width=\linewidth]{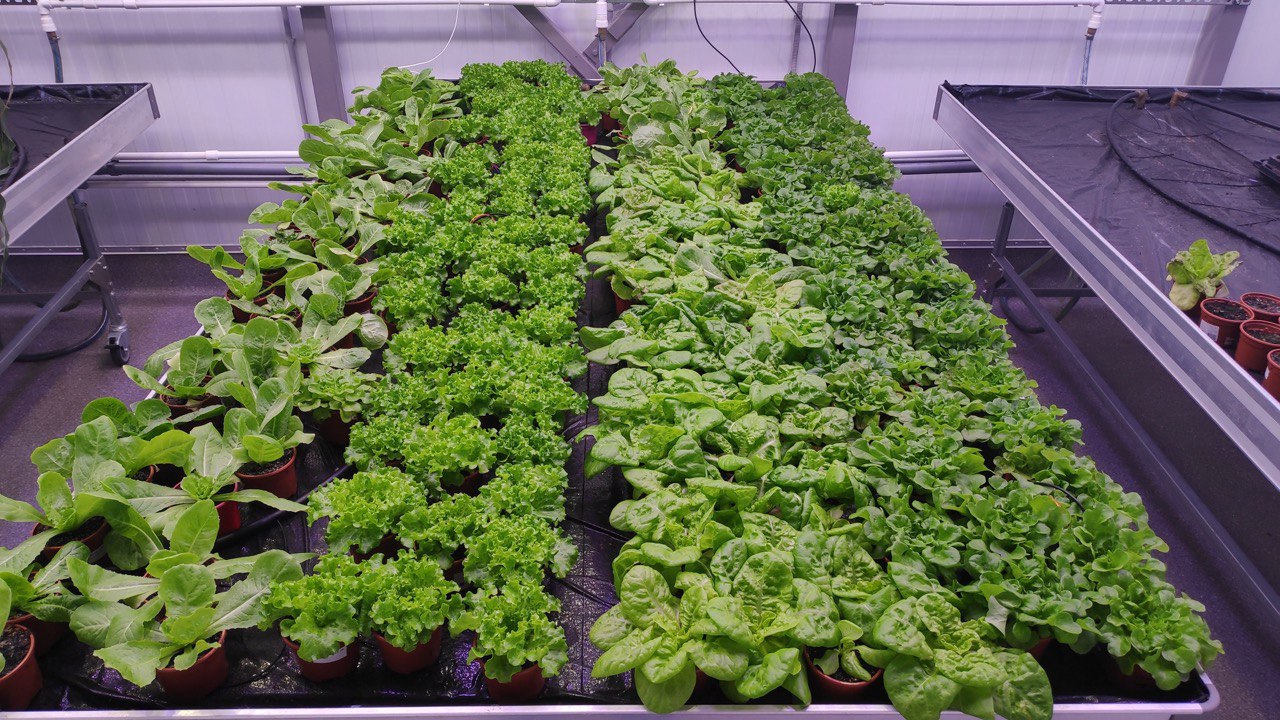}
  \caption{}
  \label{fig:setupout}
\end{subfigure}
\caption{Experimental setup. On the left side (a) the general view of the setup for EI and EII simulating a vertical farm; on the right side (b) the general view of the setup for EIII simulating an industrial greenhouse.}
\label{fig:setupphotosfull}
\end{figure}

\subsubsection{Vertical farm}
The principal characteristics of the EI and EII setup shown in Fig.\ref{fig:setupin}, are those of the designs presented in Refs.~\cite{Verticalfarm2,Lettuceyield2016Touliatos}. The setup is a vertical farm consisting of one rack fitted with four 1.5$\times$1.0 m$^2$ shelves made of galvanized steel, each separated from the next one above by a 0.5 m space. An additional shelf at the top serves as a roof for the structure. While the total ground surface area occupied by the setup is 3 m$^2$ (including 1.5 m$^2$ the surface for the watering system and a servicing area), the setup offers a total of 6 m$^2$ area available for plant growth with controllable lighting and ventilation. The vertical farm was placed inside a phytotron where the ambient humidity, air temperature, watering pump, and CO$_2$ concentration are monitored and controlled. We fixed external light-shading screens (made of multiple layers of polyethylene sheets) on the structure to leave enough space for the drainage and ventilation systems, as well as the free movement of one person, while mitigating risks of penetration of parasitic light from the surroundings into the setup. For our lighting fixture, we use SMD 2835 LEDs with a 3:1 red and blue ratio, which is claimed to be optimal for most leafy plants in several works \cite{pennisi2020optimal,azad2020evaluation}. The LED strips are evenly distributed in the bottom part of each shelf. At nominal power, they can provide a photosynthetic photon flux density (PPFD) of 145$\;\mu $mol$\cdot $sec$^{-1}\cdot $m$^{-2}$ at a  distance of 40 cm. The setup is also equipped with sensing systems for  air temperature, air humidity, CO$_2$ concentration and ambient pressure for data collection. A detailed description is given in \ref{AppA1}.

\subsubsection{Phytotron}
The EIII was performed in two identical phytotrons; one for the experimental group (DR simulation) and one for the control group. Each phytotron has independent temperature, humidity, watering control and ventilation systems. The lights used were DHlicht LED-KE300 with an evenly distributed PPFD of 220$\;\mu $mol$\cdot $sec$^{-1}\cdot $m$^{-2}$ at the soil surface. The phytotron setup is shown in Fig.~\ref{fig:setupout}. Further details on the setups, sensors, and data acquisition system the phytotron are given in \ref{AppA2}.

\subsection{Plant material and growth conditions}
Lettuce ({\it Lactuca sativa}) is one of the most widely produced leafy plants in indoor farming \cite{Verticalfarm2}. The variety {\it Lactuca sativa} (Olmetie RZ) was used as our typical reference for EI and EII. To understand the influence of genetics on light regularity sensitivity, we selected three different varieties of {\it Lactuca sativa} and one of escarole ({\it Cichorium endivia}) for EIII. The principal characteristics of each variety are shown in \ref{tab:plants}. For EI and EII, we prepared control and test groups of 100 plants respectively. For EIII we prepared control and test groups of 50 plants respectively, for each variety.

Each seed was sowed in a 9-cm diameter pot with a volume of 0.3 l where they grew during the whole experiment. To avoid damage to the roots due to transplant, the sowing depth was 1.0 cm. The pots were filled using professional-grade peat \cite{peat} and were watered during two days before planting. For the non-treated varieties(see \ref{tab:plants}), four seeds were sown per pot, and after five days, the weakest plants were thinned to one plant per pot. Later, each plant was labelled with a unique identification number. 

The plants were then placed in the experimental setup (EI and EII) and in the phytotron (EIII), where we watered them with filtered tap water three times a day every eight hours. Each time, watering was for one minute, representing 1.5 litres of water per 100 plants. The capillary mattress distributed and retained the water so that each plant got the same amount of water. Any unused water went to the drainage system. During the daytime, the ambient temperature was set to 23 $^\circ$C, and to 21 $^\circ$C for the nighttime. The relative ambient humidity was set to be $50\%$ throughout the whole day. The CO$_2$ concentration and atmospheric pressure were measured  but not controlled. Their average values were 324.6 ppm and 0.989 bar respectively.  As regards the photoperiod, during the first 15 days of growth, the plants in all groups were under the same light control profile with a 16-hour day length for EI and EII, and 11-hour day length for EIII. After day 15, we introduced lighting irregularity in the test group according to the optimized light control profile for implicit DR (discussed in section \ref{DRsimulation}).

\subsection{Light control}
\begin{figure}
    \centering
    \includegraphics[width=1\linewidth]{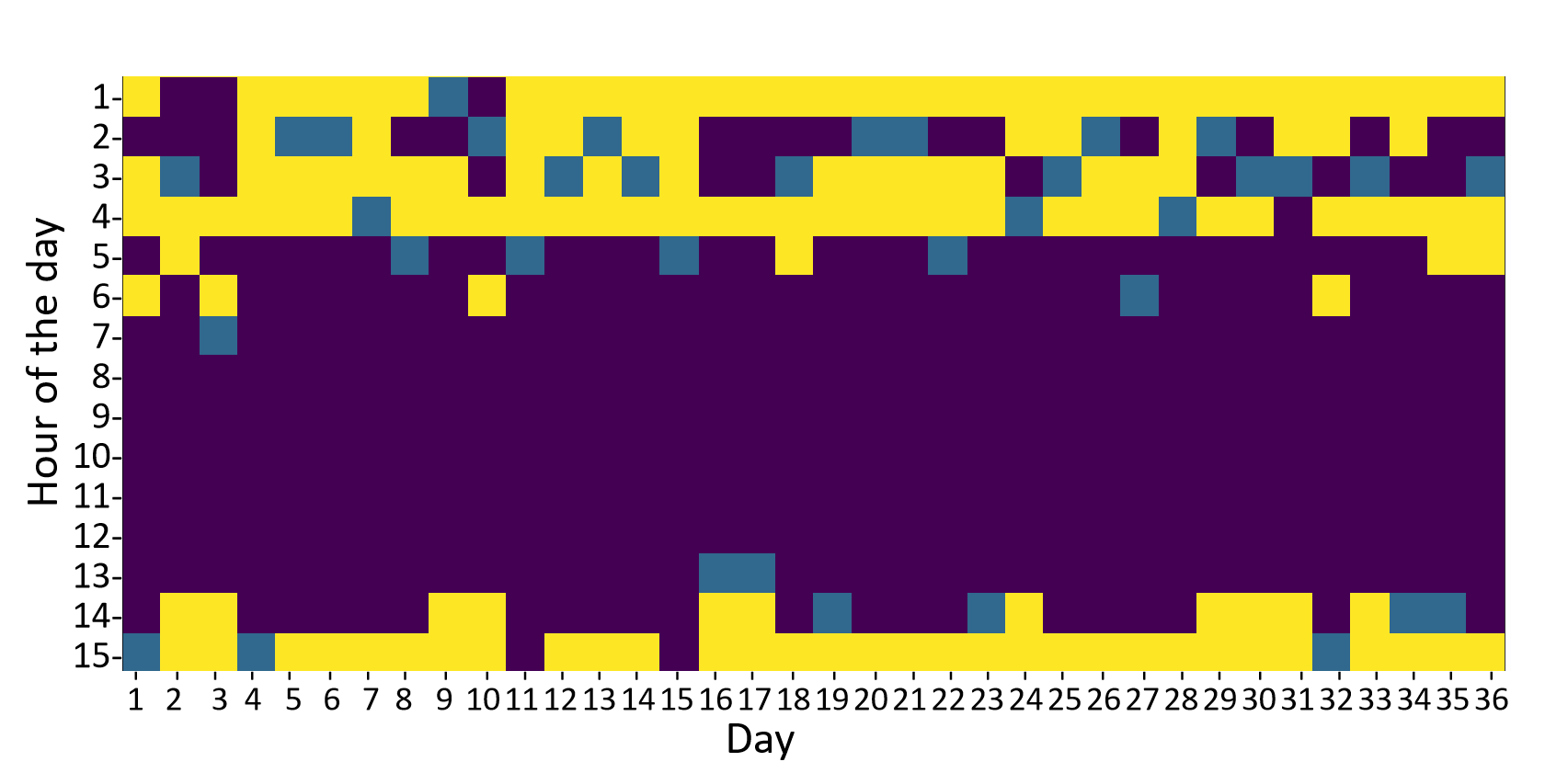}
    \caption{Illustration of the light control profile with minimized price for a selection of days from 06.07.2018 to 10.08.2018. A level of 100\% light intensity is represented in yellow, 33.33\% in green, and 0\% in dark blue. Note that from 16:00 to 0:59 we have constant 100\% light intensity.}
    \label{fig:heatmap}
\end{figure}

The light control model aims to simulate the participation of a vertical farm on implicit DR. To make the model more robust for the task, we simulated cases with extreme light irregularity (see section  \ref{DRsimulation}). A heat map representing the optimal light profile minimizing price is presented in Fig.~\ref{fig:heatmap}. The optimal outcome was adapted for convenience by shifting the daytime. Furthermore, for EIII, the photoperiod was reduced to 11 hours as the phytotron is fitted with more powerful light sources. Importantly, we kept the daily usable light, or daily lighting integral (DLI), constant across all experiments, with DLI = 8.64 mol$\cdot $m$^{-2}\cdot $day$^{-1}$, which has the lowest energy requirements as shown in \cite{Gavhane2023}. This resulted in the simplification of the thermal management of our small-scale vertical farm as we did not need to install a complex cooling system. 

\subsection{Sampling and measurements of plant physiological parameters}
The whole duration of EI was 25 days, for EII and EIII 29 days. For all experiments, during the first two weeks of growth, no samples were taken. From day 15, every five days, we randomly picked five plants from each group; and to avoid bias in the measurements due to watering, each of the selected plant was fully watered to reach saturation one hour before measuring the number of true leaves, the fresh mass and the dry mass including the root system. 

The fresh weight of the plants were measured using an 
electronic balance with a resolution of 0.01 g. To measure the dry mass of the plants, we packed them in separate open bond paper envelopes and put them to dry in a 
drying oven with forced convection, for 72 hours at 80$^\circ$C with full ventilation power (4800 l/h). After the plants had dried, we used an 
electronic balance with a resolution of $10^{-4}$ g to measure the dry mass.

\subsection{Environmental parameters measurements}
From the first day of the experiment, the indoor microclimate parameters (temperature, relative humidity, atmospheric pressure, carbon dioxide concentration) were measured every 4 minutes. The PPFD was measured using a quantum sensor at the beginning and the end of the experiments. That showed that the lighting system did not degrade during the whole course of the experiments.

\subsection{Demand response simulation}
\label{DRsimulation}
Using the spot energy price data available on the website of the Russian energy market administrator \cite{DRdata}, we simulated the lighting of a vertical farm participating in implicit demand response. The available data is from 2017 to the current date; the only typical year in that range was 2018. A typical year is understood here as a year when: 
\begin{itemize}
    \item The spot energy price at the beginning and the end of the year are similar;
    \item No tax or regulation-related changes were introduced;
    \item No major atypical event such as, e.g., economic crisis, pandemics or conflict, affects the consumption profiles or the infrastructure;
    \item No significant natural phenomenon such as heat wave or harsh cold weather affects the consumption profile or the infrastructure.
\end{itemize}

\begin{figure}
    \centering
    \includegraphics[scale=0.9]{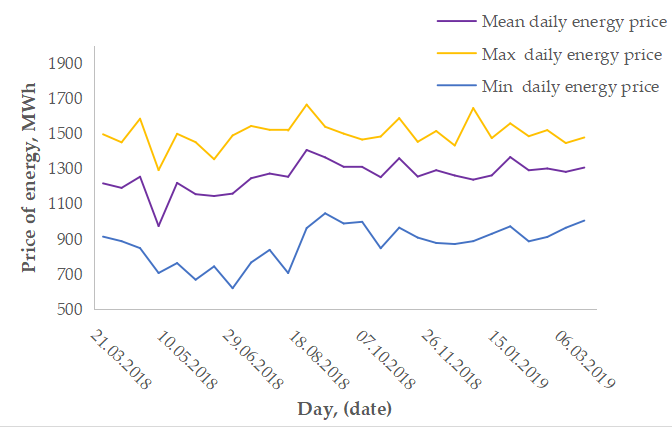}
    \caption{Daily maximum, minimum and average electricity spot prices in 2018.}
    \label{fig:DR2018absolutev}
\end{figure}

The whole year's data was analyzed to find the time of the year when the spot price of electricity experiences the most significant daily changes. We defined the most variable days as those with the largest difference between their maximum and minimum daily price. The daily mean, minimum, and maximum spot prices are shown in fig. \ref{fig:DR2018absolutev}. As the most significant differences in hourly electricity spot prices are observed in the middle of July, we selected a period of $\pm$ 25 days centered on the 15th of July for our study. The data is used for a linear optimization problem to find the optimal light control profile by minimizing the overall daily energy cost. The minimization problem is solved for each day separately, as the light needs of the plants are assumed to be not transferable to other days.

The most robust simulation test we performed was under the condition of the most irregular lighting profile selected for our simulations. Here, most irregular profile means a sequence with the largest number of ``day/night'' (LEDs switched on and off) cycles per day, with the most irregular on/off duration pattern. We used the Python library Scypy to solve the linear optimization problem formulated as 

\begin{equation} \label{eq:1}
\frac{A}{k_{\rm ph}}~~{\rm min} \sum_{1}^{24} C_{i}~ \eta_{i}
\end{equation}

\noindent subject to:

\begin{equation}\label{eq:2}
\sum_{1}^{24} \eta_{i}=R_{\rm opt} \end{equation}
\begin{equation}\label{eq:3}
0\leq \eta_{i}\leq R_{\rm opt}/h \end{equation}

\noindent where A is the total illuminated area, $C_i$ and $\eta_i $ are respectively the price of a MWh of electric energy and the hourly PPFD at a given hour $i$, $R_{\rm opt}$ is the optimal DLI, $h$ is the duration of the daylight, and $k_{\rm ph}$ is the measured photon efficacy \cite{LEDefficacy} of our lighting fixture. The resulting light control profile, depicted in Fig.~\ref{fig:heatmap}, was found under the constraints of $R_{\rm opt}=14.4$ mol$\cdot $m$^{-2}\cdot $day$^{-1}$ and $h=14$ hours. The measured photon efficacy  $k_{\rm ph}$ \cite{LEDefficacy} of our lighting fixture, reads:

\begin{equation}\label{eq:4}
k_{\rm ph}=\frac{{\rm PPFD}\times A}{P_{\rm cons}} =\frac{150\times 3.6}{350} = 1.7~\mu {\rm mol}\cdot {\rm J}^{-1}
\end{equation}

\noindent and $A$ was determined assuming a uniform distribution of LEDs for each shelf and a beam angle of 90$^{\circ}$; $P_{\rm cons}$ is the measured averaged power consumption of the LEDs.

The validation of our calculation was simply performed by comparison of the value we found with that indicated in~\cite{LEDefficacy}. The experimentally determined efficacy in our setup is about 2/3 of the lowest value for a blue/red fixture reported in \cite{LEDefficacy}. This could be expected as ensuring the maximum efficacy conditions as those in \cite{LEDefficacy} was not an objective in the frame of the present work.

\subsection{Statistical data analysis}
Statistical analysis was performed with two-way ANOVA analysis. To ensure the independence of observations across the different times, plants were randomly selected for the measurements. As the experimental data reflects the accelerated growth of the plants, data were transformed by using the logarithm of the numerical data value \cite{mcdonald2009handbook}. This allows to increase the homogeneity of the variances and the statistical distributions of the residuals of ANOVA models become close to the normal distribution. The measured plant growth parameters fresh biomass and dry biomass are reported as their $\log_{10}$ value and the quantity of true leaves as their natural logarithm, $\ln$, value. All computations were done using R software ver. 4.3.

We performed the ANOVA analysis for EI and EII together (as EII is a replica of EI) and for EIII separately. The model used for EI and EII is:

\begin{equation}\label{eq:AnovaEI_EII}
y_{ijkl} = \alpha_i \times \beta_j + \gamma_k + \epsilon_{ijkl}
\end{equation}
\noindent with $y_{ijkl}$ being the recorded trait for each plant ( fresh and dry biomass, number of leaves) appropriately transformed, $\alpha$ the fixed factor for time, $\beta$ the fixed factor for experiment, $\gamma$ the fixed factor for condition and $\epsilon_{ijkl}$ the residuals of the model with expected $iid\quad \mathcal{N}(0, \sigma^2)$ distribution.

The model used for EIII reads:
\begin{equation}\label{eq:AnovaEIII}
y_{ijkl} = \alpha_i \times \beta_j \times \gamma_k + \epsilon_{ijkl}
\end{equation}

\noindent where, similarly, $y_{ijkl}$ is the recorded trait ( fresh and dry biomass, number of leaves) appropriately transformed, $\alpha$ the fixed factor for time, $\beta$ the fixed factor for experiment, $\gamma$ the fixed factor for condition and $\epsilon_{ijkl}$ the residuals of the model with expected $iid\quad \mathcal{N}(0, \sigma^2)$ distribution. 

In Eqs.~(\ref{eq:AnovaEI_EII}) and (\ref{eq:AnovaEIII}), the sum $+$ indicates that the predictors have only additive effect: moving up or down the regression line without changing the slope, as they do not interact with other predictors; the product $\times$ indicates that a predictor interacts with other predictors, meaning that it has both an additive effect and an effect that depends on the other predictors with which it interacts.

\section{Results}
\label{res}

\subsection{Data collection}
For each experiment, the fresh weight, dry weight, and quantity of true leaves were collected as described in the Material \& Methods section. The microclimate parameters were collected during all the experiments. An example of the raw data can be found in Fig. \ref{fig:Sample_data}. We calculated the average values of the environmental parameters. An example of the collected data set can be found in Table \ref{tab:PlantData}. Note that for plant growth analyses, from day 0 (sowing) to day 15, all plants were growing under the same conditions with classical light regime.

\begin{figure}
    \centering
    \includegraphics[width=\textwidth]{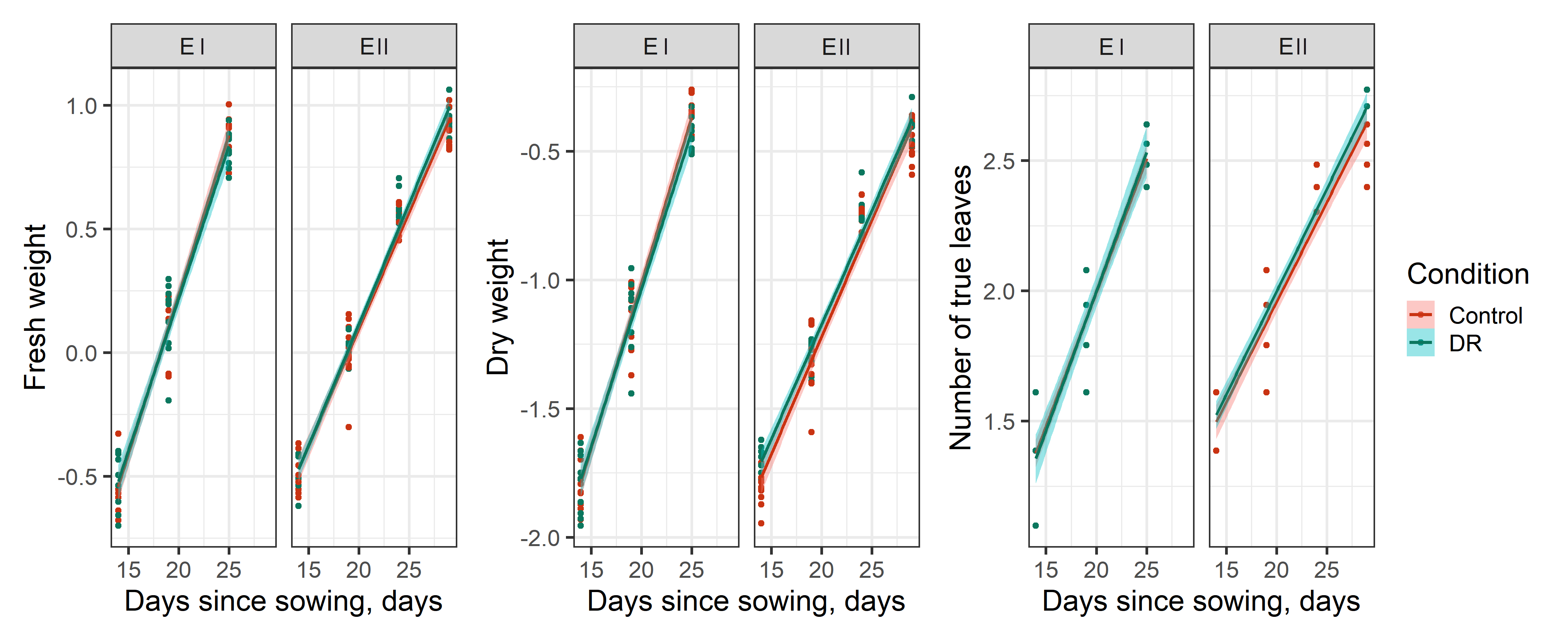}
    \caption{Comparison of growth curves in replicates EI and EII for DR (in blue) and control (red) condition groups. In the left panel, the measured fresh weight is normalized to 1 g and reported on a $\log_{10}$ scale; in the middle panel, the measured dry weight is normalized to 1 g and reported on $\log_{10}$ scale; and in the right panel, the number of true leaves is reported on a $\ln$ scale. The blue and red lines correspond respectively to the predicted values for control and demand-response from the linear model of equation \ref{eq:AnovaEI_EII}. No statistically significant differences were found due to condition, i.e. control or demand response (Table \ref{tab:ANOVA}) by the two-way ANOVA.}
    \label{fig:EI_EII_results}
\end{figure}

A first replicated experiment (EI and EII) to compare the plant production under demand response (DR) and standard (control) condition was performed using the Olmetia lettuce variety, during 29 days. The increase of fresh biomass, dry biomass and number of leaves as a function of time is displayed in Figure \ref{fig:EI_EII_results}. A second set of experiments (EIII) analysing the response of the three different traits for four different green vegetable varieties in response to demand-response or classical regime was also performed during 30 days. The increase of fresh biomass, dry biomass and number of leaves as a function of time for each variety in the two lightning conditions is shown in Figure \ref{fig:EIII_results}. For EIII, demand-response and classical light regime were set-up in different phytotrons.

\subsection{Statistical data analysis}

\begin{figure}
    \centering
    \includegraphics[width=\textwidth]{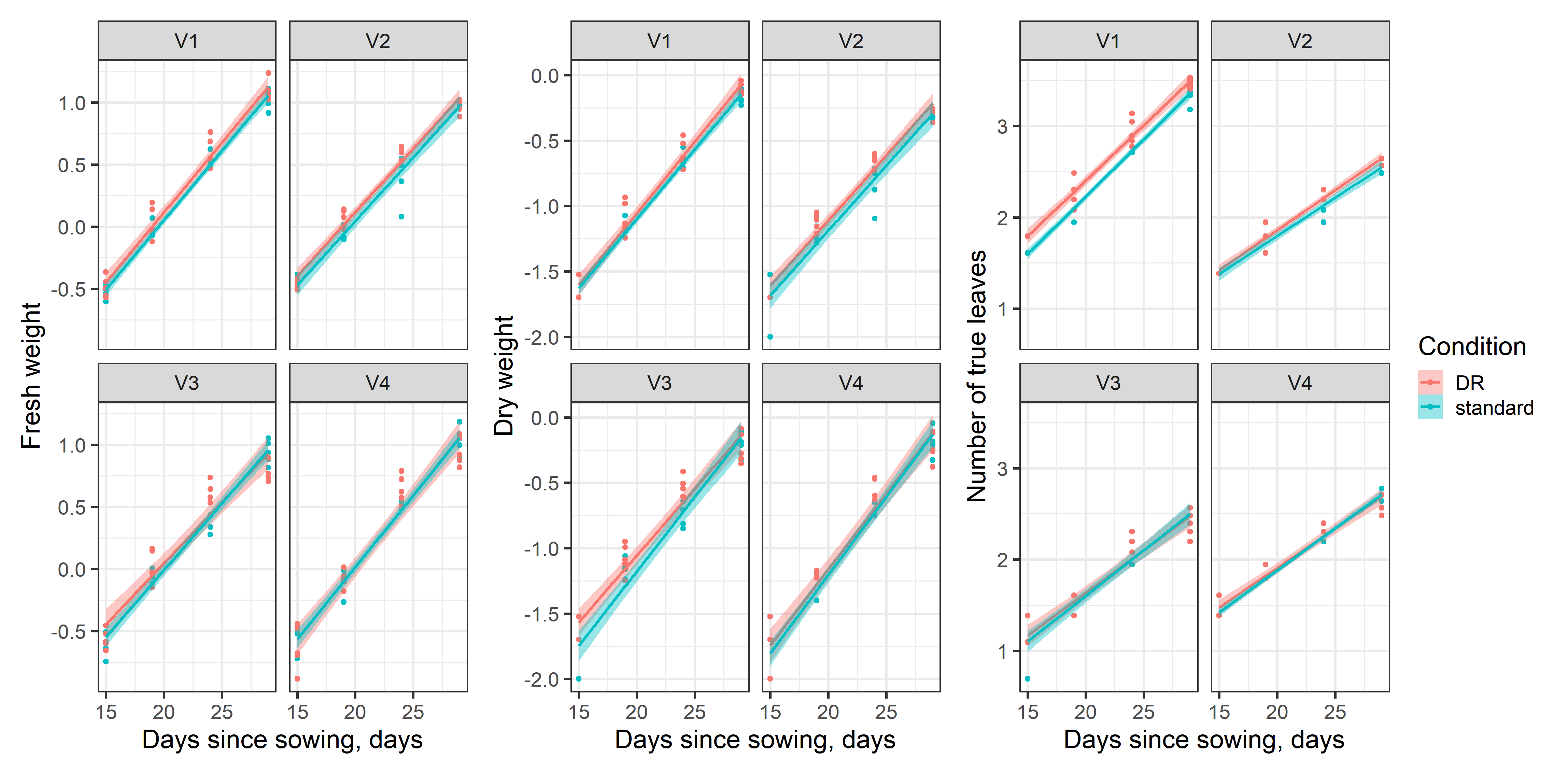}
    \caption{Comparison of growth curves depending on DR and control conditions for four different plant varieties (V1,V2,V3 and V4). In the left panel, the measured fresh weights are reported normalized to 1g in $\log_{10}$ scale; in the middle panel, the measured dry weights are reported normalized to 1g in $\log_{10}$ scale; in the right panel, the number of true leaves is reported in $\ln$ scale. For each plot, the blue and red lines correspond respectively to the predicted values for control and demand-response from the linear model of Eq.~\ref{eq:AnovaEIII}. Statistically significant differences ($p\leq0.05$) were found with the two-way ANOVA test due to condition and variety. The interaction of time (number of days since sowing) and condition (DR or control) were  statistically insignificant, so condition has no effect on the growth rate, and only genetic variability affects growth rate.}
    \label{fig:EIII_results}
\end{figure}

The outcome of ANOVA analyses (summarized in Table \ref{tab:ANOVA} in \ref{app:Data}) shows that growth under DR lighting regime or under standard day length, \textit{i.e.} the condition factor, has no effect on any of the recorded traits for EI and EII. Even if the growth rate was different between the two experiments (significant p-values for Days $\times$ Experiment factor), this was not affected by the condition. 
 
For EIII, besides the obvious effect of Days on the increase of the traits, we found differences due  to Variety and Condition. This latter effect is an offset, indicating that at the start of the light regime experiment - after 15 days -  the plants have been previously growing differently in the phytotrons under standard light regime. We thus see that the phytotron that produced bigger plants was then affected to the demand-response regime. 
 
Actually, we are looking for a significant interaction between Days and Condition, that would indicate that the growth rate is affected by the light regime; and possibly by a complex Days $\times$ Condition $\times$ Variety interaction that would suggest that the light regime affects the growth of plants in a variety-dependant manner.  We demonstrate that the growth rate of plants depends on the variety due to significant interaction between Variety and Days effect. However, non-significant interactions between time and condition were revealed (Table \ref{tab:ANOVA} and Figure \ref{fig:EIII_results}, thereby implying that the light regime does not affect the growth rate for several parameters of the four assessed varieties.

\section{Evaluation of potential economic impact}
\label{disc}
\subsection{Hypotheses}
Having evidenced that there is no significant impact of DR participation on the leafy plants production, we may turn to the evaluation of the potential economic benefits for the grid and for the indoor farming industry. As we do not have access to control and sensing data from the industry, we can only provide an evaluation of orders of magnitude because of insufficient information. We thus keep our techno-economic analysis simple to try and draw meaningful conclusions that may easily be tested. We start by simulating the base case using the energy price data of every 15-th day of each month \cite{DRdata} and the solar radiation profile of our model year 2018. Then we proceed with implicit DR participation, and finally calculate the benefits for explicit DR participation. Our numerical simulations are based on the following assumptions:

\begin{itemize}
    \item Since in Russia there are mostly greenhouses and comparatively much less vertical farms \cite{Ivanov2018modernization}, we model the whole load as an aggregation of greenhouses;
    \item Co-generation in indoor farming is a known practice \cite{Ivanov2018modernization,minitets,2015biogas,agroinvestor_2021}. However, there is no available information on the size of this practice. We will assume that the whole industry is connected to the grid and has no local production;
    \item We restrict our analysis to the needs of the load for complementary lighting;
   \item From Meteonorm \cite{meteonorm} we obtained the measured photosynthetic active radiation for every day of our model year 2018. This measurement is a ground measurement (Moscow, Sheremetyevo airport) meaning it takes into account variations in cloudiness and other environmental factors; so we assume that all weather and climate related phenomena are already accounted for in the data used in the present work.
    \item The simulated greenhouse aims to match the maximum DLI from solar radiation, which for Moscow is 14 mol$\cdot $m$^{-2}\cdot$day$^{-1}$ (see table \ref{tab:PARtable}), without supplementary CO$_2$ feeding;
    \item The greenhouses' walls solar radiation transmission coefficient is assumed to be equal to 0.8 \cite{vanDelden2021,Verticalfarm2,sanford2011reducing};
    \item The total incoming radiation on plants in greenhouses is kept stable and compensates for the lack of solar radiation using complementary lighting. The lamps are assumed to be high-pressure sodium (HPS) lamps with a maximum PPFD from 250 $\mu$mol$\cdot $m$^{-2}\cdot$s$^{-1}$  and an efficacy  of 1.7 $\mu$mol$\cdot $J$^{-1}$ (identical to our small-scale vertical farm). The selected efficacy is slightly bellow the best tested HPS lamps in \cite{LEDefficacy}. 
\end{itemize}

\subsection{Electricity consumption profile model}
We calculate the monthly average of hourly PAR measurements accounting for the transparency of the greenhouse walls as follows:

\begin{equation} \label{eq:5}
\rho _i= \frac{3600~ k_{\rm PAR}~ K_{\rm trans}}{N}~\sum_{j=1}^{N} R_i^j
\end{equation} 

\noindent here $N$ is the number of days per month, $\rho_i$ is the average hourly PAR for a given month inside of the greenhouse, $R_i$ is the hourly measured PAR, $ k_{\rm PAR}=2.02 $ is the conversion factor \cite{Foken2008, Mavi2004}, $K_{\rm trans}=0.8$ is the light transmission coefficient of the greenhouse walls. The average daily radiation per month is given in Table~\ref{tab:PARtable}.

\begin{table}[]
\centering
\begin{tabular}{|l|l|l|l|l|l|}
\hline
\hline
\textbf{January}  & \textbf{February} & \textbf{March}     & \textbf{April}    \\ 
1.28              & 3.01              & 6.06               & 9.20             \\ \hline            
 \textbf{May}      & \textbf{June} & \textbf{July}     & \textbf{August}  \\ 
12.47             & 14.00 & 13.69             & 10.93             \\ \hline
 \textbf{September} & \textbf{October}  & \textbf{November} & \textbf{December} \\ 
   6.8               & 3.43              & 1.40              & 0.87  \\ \hline\hline     
\end{tabular}
\caption{Calculated average daily photosynthetic active radiation in mol$\cdot$m$^{-2}\cdot$day$^{-1}$, for every month inside a greenhouse in the Moscow region.}
\label{tab:PARtable}
\end{table}
 To calculate the impact on the grid, we determine first the greenhouse consumption profile (base case). Using the assumptions we made before, we solve the problem as an optimization problem to minimize the variability of light radiation during the day:
 
\begin{equation}\label{eq:6}
{\rm min} \sum_{1}^{24} -\omega_i~\eta_{i}\end{equation}

Subject to:
\begin{equation}\label{eq:7}
\sum_{1}^{24} \eta_{i}=R_{\rm opt} - \sum_{1}^{24}\rho_{i} \end{equation}
\begin{equation}\label{eq:8}
0\leq \eta_{i}\leq \frac{R_{\rm opt}}{h} -\rho_{i} \end{equation}

\noindent where $\omega_i$ is the uniformity criterion centered on the day time 12 pm with the integer $1\leq i\leq 24$. We choose the form $\omega_i = \sin(i\pi/24)$ as it has the two other following suitable properties: $0\leq \omega \leq 1$ and symmetry: the further the time is away from 12 pm, the smaller $\omega_i$ is. Note that as we sample the sine function, the optimization problem remains linear. The quantity $\rho_{i}$ is clipped  for the inequality constraint so that $\frac{R_{\rm opt}}{h} -\rho_{i} \geq 0$, and we always have a feasibility region for every value of $\eta_{i}$. The resulting lighting profile is multiplied by the number of days per month $N$, the ratio $A/k_{\rm ph}$ with  $A = 10000$ m$^2$, and the hourly energy cost for every 15-th day of each month, $C_i$ to discover the base case energy cost for a hectare per month and per year, the results are shown in Tables \ref{tab:savingspermonth} and \ref{tab:savings} respectively.

\subsection{Economical impact of DR on the indoor farming industry}
\subsubsection{Implicit DR}
\begin{figure}
    \centering
    \includegraphics[scale=0.55]{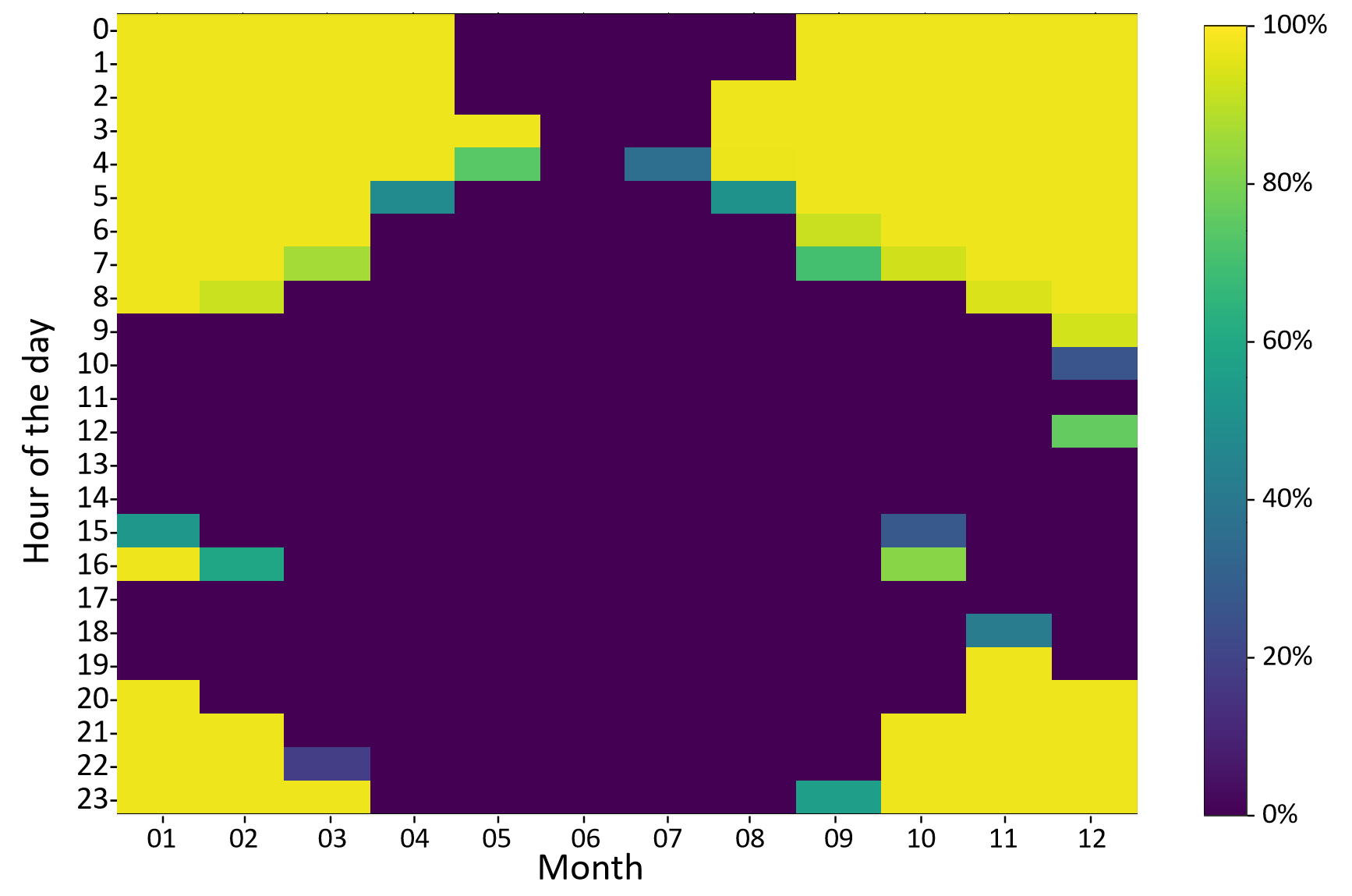}
    \caption{Heat map showing the optimized  complementary lighting profile (optimal implicit DR participation)  for the 15-th of each month.}
    \label{fig:heatmapopt}
\end{figure}

The economic benefit of different DR programs for the indoor farming sector is calculated as follows. For the implicit DR, we use a linear minimization problem to simulate a participation similar to the base scenario: we minimize Eq.~(\ref{eq:1}), subject to Eq.~(\ref{eq:7}) and Eq.~(\ref{eq:8}) where $C_i$ is the hourly energy cost for every 15-th day of each month. The optimized consumption profile for implicit DR is shown in Fig.~\ref{fig:heatmapopt}. The current estimation does not consider restrictions for the number of night hours, which are affected differently depending on the season. The resulting profile is then multiplied by the number of days per month $N$. The overall costs of electricity for the implicit DR case for a hectare per month is presented in Table \ref{tab:savingspermonth}, and the total costs as well as the calculated savings are presented in Table \ref{tab:savings}.

\subsubsection{Explicit DR}
\begin{figure}
    \centering
    \includegraphics[scale=0.55]{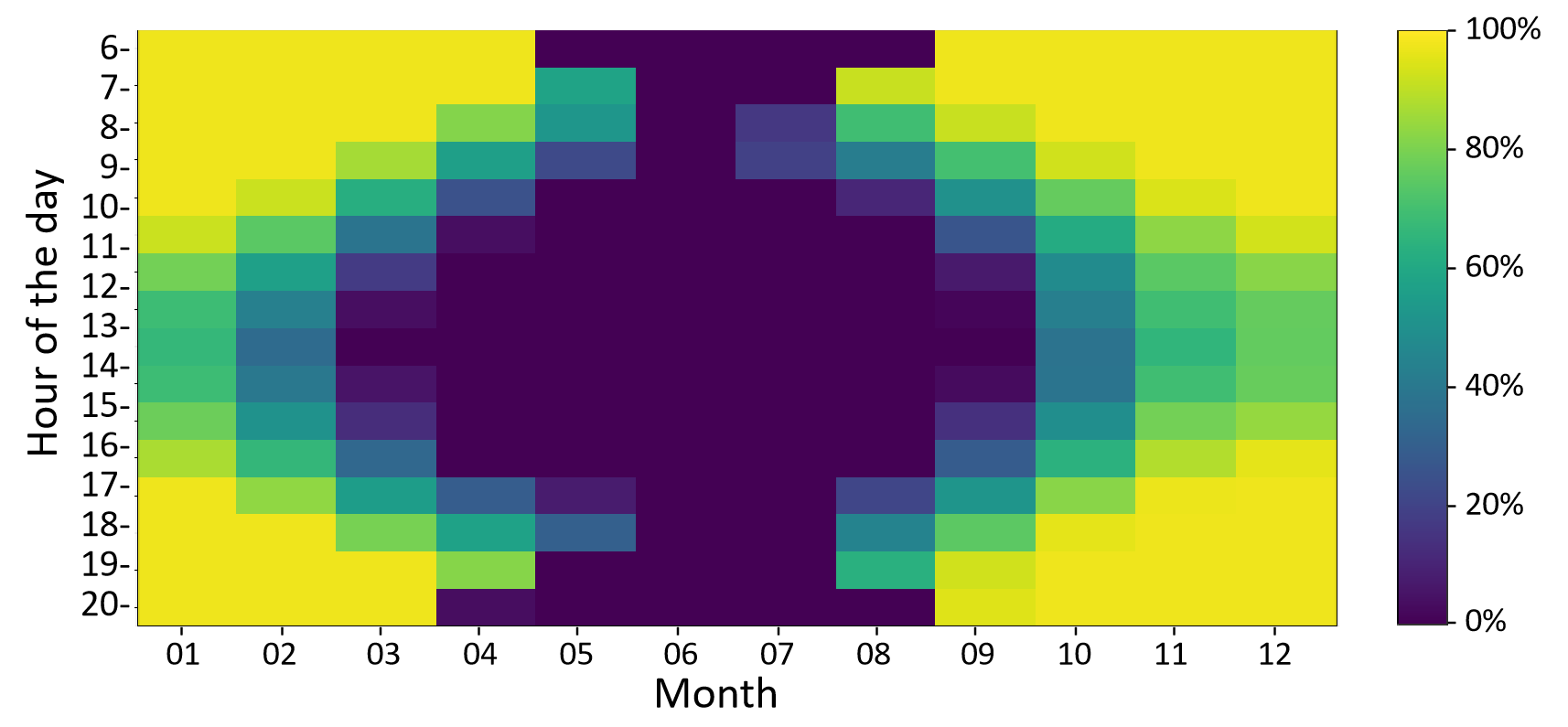}
    \caption{Heat map showing the estimated complementary lighting profile on the 15-th of each month. From 20:00 to 6:00 there can be no complementary lighting, i.e. only natural night depending on the season.}
    \label{fig:heatmapcurrent}
\end{figure}

The data reported in Table~\ref{tab:savingspermonth} shows that from May to August participation in explicit DR presents no interest as the load consumption level is too low; hence quarters Q2 and Q3 of the year are included in the analysis only in a mixed implicit-explicit DR strategy (as the DR participation contract is for three months \cite{VygonDR2018,energynet2019}). The greenhouses can participate in explicit DR during the Q1 and Q4 quarters of the year. Here, the maximum participation is calculated as the minimum hourly load during peak time (from 8 am to 9 pm) \cite{Peak_2023} for the participating 3-month period. The calculated minimum load for DR participation is 0.49 MW $\cdot$ha$^{-1}$ for Q1 and 0.55 MW $\cdot$ha$^{-1}$ for Q4. Note that during the month of March, as shown in Fig.~\ref{fig:heatmapcurrent}, there is a risk that the load is not available upon the request of the energy system operator for DR participation. This may result in a financial penalty on the load side; so, assuming that the penalty would cancel the benefits for the month of March, we scale-down the Q1 benefits to 2/3 of what they would be if the load could be available for participation at any requested time. To estimate the economic benefit, we use the average monthly payment for the first pricing zone in the Russian DR program, which during 2020 was 346,582.44 rub$\cdot $MW$^{-1}\cdot$month$^{-1}$\cite{systems_2021,electricity_2021}. The total amount to be paid depends on the initial bid, the base load to disconnect, the successful demonstration of readiness, and the successful demonstration of participation in the peak shaving events \cite{yyear_2021}. 

Assuming the readiness and peak shaving events are successfully demonstrated, we can calculate the yearly economic benefits for the Russian indoor farming industry per hectare. Further, we can add up the economic benefits from implicit DR for Q2 and Q3 to explicit DR for a mixed implicit-explicit DR strategy. The results of the calculation are presented in Table~\ref{tab:savings}. Currently, the overall area used for indoor farming in Russia is around 3372 ha. The largest share of cultivated area is for cucumbers with 55.3\%, followed by tomatoes with 41.7\%, which leaves 3\% for leafy plants \cite{agroinvestor_2021, IKAR2023}. 

\subsubsection{Effect of the total cultivation surface on the economic impact}
Let us consider two simple scenarios of vegetable production. On scenario - ScI - involves only leafy plants, for which we demonstrate that DR does not affect production. The second scenario - ScII - involves the production of cucumber and tomato, both vegetables known to be neutral to day-length, meaning they can flower under short-night/long-day or long-night/short-day conditions, therefore likely not affected by a DR-based light regime. For ScI, only leafy plants growers, representing 3\% of the indoor farming industry, participate in DR. For ScII, we include the acreage destined for the vegetative growth stage of cucumber (approximately 30 out of 110 days, for a total of 27.3\% of total cultivation time \cite{cucumber2017}) and tomatoes (approximately 45 out of 121 days -- including three weeks for 100\%  harvest, for a total of 37.1\% of total cultivation time \cite{OECD2017}), resulting in 30.6\% of the total cultivation time. Hence, for the scenario ScII, this amounts to assuming that the percentage of the total cultivation time is equivalent to the percentage of acreage used also adding that used for green-leafy plants, for a total of 33.6\% of the total land used. Note that the absence of negative effects for plants growing under lighting constraints imposed by implicit DR participation (and by extension explicit DR too) demonstrated in section \ref{res} is valid for the vegetative growth stage of leafy vegetables. In the second scenario (ScII), we assume that there is no negative effects of DR, thanks to the day-neutral flowering habit of tomato and cucumber. Hence, for ScI the total load that could participate in DR in 2022 was 49.57 MW for quarter Q1 and 55.6 MW for Q4. For ScII the total load that could participate in DR in 2022 was 555.2 MW for quarter Q1 and 623.1 MW for Q4. The total saving for both scenarios are shown in Table \ref{tab:savings}.

\begin{table}[]
\begin{tabular}{|l|rrrr|rrr|}
\hline
\multicolumn{1}{|c|}{\multirow{2}{*}{\textbf{Case}}} &
  \multicolumn{1}{c|}{{\textbf{Cost}}} &
  \multicolumn{3}{c|}{\textbf{Savings industry side}} &
  \multicolumn{3}{c|}{\textbf{Savings grid side}} \\ 
\multicolumn{1}{|c|}{} &
  \multicolumn{1}{c|}{\textbf{per ha}} &
  \multicolumn{1}{c|}{\textbf{per ha}} &
  \multicolumn{1}{c|}{\textbf{\begin{tabular}[c]{@{}c@{}}ScI \\ \end{tabular}}} &
  \multicolumn{1}{c|}{\textbf{\begin{tabular}[c]{@{}c@{}}ScII \\ \end{tabular}}} &
  \multicolumn{1}{c|}{\textbf{per ha}} &
  \multicolumn{1}{c|}{\textbf{\begin{tabular}[c]{@{}c@{}}ScI \\ \end{tabular}}} &
  \multicolumn{1}{c|}{\textbf{\begin{tabular}[c]{@{}c@{}}ScII \\ \end{tabular}}} \\ \hline

\textbf{\begin{tabular}[c]{@{}l@{}}Implicit\\ DR\end{tabular}} &
  \multicolumn{1}{r|}{4.71} &
  \multicolumn{1}{r|}{0.85} &
  \multicolumn{1}{r|}{86.0} & 
  \multicolumn{1}{r|}{963.0}
   &
  \multicolumn{1}{r|}{5.19} &
  \multicolumn{1}{r|}{525.0} &
   \multicolumn{1}{r|}{5880.2}
   \\ \hline
\textbf{\begin{tabular}[c]{@{}l@{}}Explicit \\ DR\end{tabular}} &
  \multicolumn{1}{r|}{4.48} &
  \multicolumn{1}{r|}{1.08} &
  \multicolumn{1}{r|}{109.3} &
   \multicolumn{1}{r|}{1223.6}
   &
  \multicolumn{1}{r|}{5.28} &
  \multicolumn{1}{r|}{534.1} &
   \multicolumn{1}{r|}{5982.2}
   \\ \hline
\textbf{\begin{tabular}[c]{@{}l@{}}Mixed \\ strategy\end{tabular}} &
  \multicolumn{1}{r|}{4.28} &
  \multicolumn{1}{r|}{1.28} &
  \multicolumn{1}{r|}{129.5} &
  \multicolumn{1}{r|}{1450.2}
   &
  \multicolumn{1}{r|}{8.81} &
  \multicolumn{1}{r|}{891.2} &
  \multicolumn{1}{r|}{9981.7}
   \\ \hline
\end{tabular}
\caption{Results of the economic assessment (in Mrub/year) of the participation of implicit DR, explicit DR and a mix strategy implicit DR Q2 and Q3 + explicit DR Q1 and Q4, considering the ScI and ScII scenarios. As of 2024 the USD/RUB conversion rate is assumed to be around 1/90.The calculated base case cost is 5.56  MRub/year$\cdot$ha.}
\label{tab:savings}
\end{table}

For explicit DR, the estimated power generation that can be displaced with the current DR program that holds a maximum of five peak shaving events per month is currently in the 6,200 MW range \cite{energynet2019}, which is much larger than the Russian indoor farming aggregated load at any time of the year (the calculated peak power demand being 1.4 MW $\cdot $ha$^{-1}$). Therefore, the size of the Russian DR program does not represent a limitation for any of our studied scenarios. 

\subsection{Economic impact of DR on the grid side}
The economic impact estimation of DR on the grid operation is a quite complicated task and an approach to calculate it is still an object of discussion and research \cite{energynet2019,GOOD2019107}. While the development of a full model is beyond the scope of the present work, we evaluate the economic impact of explicit DR in the Russian Federation relying on the estimates published by the Russian national initiative EnergyNet \cite{energynet2019}. Explicit DR has an estimated economic impact of 105 billion rub$\cdot$year$^{-1}$ for a maximum participation of 6200 MW \cite{energynet2019}. Assuming a linear correlation between the overall economic impact and the total participation, the economic benefit for the grid is 16.95 million rub$\cdot$year$^{-1}\cdot$MW$^{-1}$. For the explicit DR and implicit DR cases, some of the benefits in infrastructure cost reduction are not met, as during the quarters Q2 and Q3, electricity is consumed during several hours at peak time to meet the needs of the load. As shown in \cite{energynet2019}, about 40\% of the positive economic effects of DR is expected in infrastructure cost reduction, so the overall economic benefit is reduced proportionally. For peak shifting the participating load is calculated as the yearly average displaced load from peak time. In Russia, because of the size of the country, the accumulated daily peak time duration is on average 12.16 hours per month in Russia \cite{Peak_2023}. The economic impact of peak shaving and load shifting holds for the grid in the mixed implicit-explicit DR strategy as the peak time load is reduced to the maximum base load during explicit DR participation and to zero during implicit DR participation. The resulting savings are shown in Table~\ref{tab:savings}.

\subsection{Carbon footprint analysis}
While peaking power plants, including hydroelectric and gas or coal-fired power plants are the most controllable electricity generation units \cite{Generation_2023}, they operate at sub-optimal technical capabilities most of the time due to the capacity reserved for peaking, hence resulting in an increased use of fossil fuel \cite{ma2022multi, ebrahimi2014combined, ekwonu2013modelling}. If the need for peaking power plants and the related infrastructure could be significantly reduced by DR programs, fossil fuel consumption would be reduced, thus allowing the energy system to operate existing power plants in high-efficiency regimes and possibly fostering further penetration of solar and wind energy sources \cite{siano2014demand, iea_2020}. So, one may anticipate that by decreasing its dependence on the electricity produced by peaking power-plants, the indoor agriculture industry's carbon footprint can be reduced. This hypothesis should be examined.

\begin{figure}
    \centering
    \includegraphics[width=\linewidth]{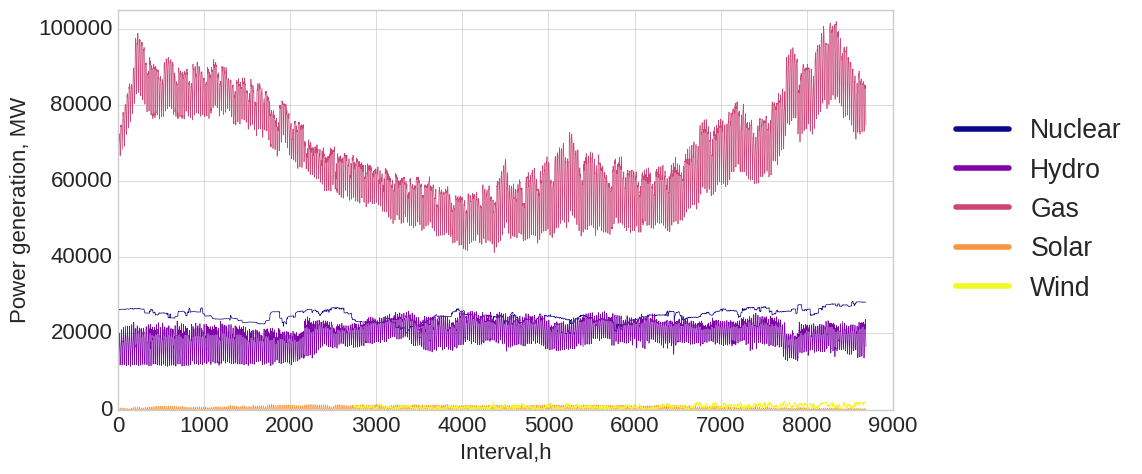}
    \caption{Mix of electric power generation for the Russian unified energy system in 2023.}
    \label{fig:GenerationMix}
\end{figure}

An accurate evaluation of the carbon footprint reduction assuming the full adoption of DR in the Russian indoor farming industry, necessitates a full life cycle assessment for the Russian energy market. Limiting ourselves to the publicly available information, we approximate the carbon footprint reduction using the results of the 2022 Carbon Neutrality in the UNECE Region: Integrated Life-cycle Assessment of Electricity Sources report \cite{gibon2022carbon}. From the mix of electrical power generation of the Russian unified energy system in 2023 \cite{Generation_2023} depicted in Fig. \ref{fig:GenerationMix}, we determine that only thermal and hydroelectric power stations participate in the spot market, with the average daily variability for both types of power-plants and the participation ratio on peaking generation being 64\% thermal and 36\% hydroelectric power-plants. Then, using the overall global average footprint per kilowatt hour of electrical energy produced reported by the IPCC \cite{gibon2022carbon}, we estimate the carbon intensity of electricity production during peak time to be 322 grams of CO$_2$-equivalent per KWh of electric energy. This result is quite close to the average carbon intensity of electricity production found in the literature \cite{veselov2022,owid-electricity-mix}. Therefore, the carbon footprint reduction due to participation in implicit DR is negligible in the Russian Federation, as carbon intensity of energy generation during peak time is equivalent to the average carbon intensity. Further, as our experimental results shown in Section~\ref{res} suggest, the plants mass increases during the vegetative growth in green leafy plants is proportional to the energy input, meaning that to reach the target weight for a vegetable to be sold, producers participating in explicit DR, should compensate the peak shaving participation hours with additional hours of lighting. In other words, the participation in explicit DR does not reduce the overall energy consumption, but redistributes it. Hence, similarly to the case of peak shifting, the effects in carbon emission are negligible. 

The counter-intuitive result that an environmentally friendly technology as DR may not reduce the carbon footprint of the electricity consumed can be attributed to two different problems: one is the lack of data on the distribution of the energy mix by fuel, which precludes precise estimations; the other is the absence of a constraining emissions reduction policy, which leads to coal-fired power-plants competing with gas-fired power-plants, more efficient convened cycle power-plants, and hydroelectric power-plants under equal conditions. Introduction of carbon taxes, segregation from base load generation using policies or subsidies for more efficient power production could give DR the capability to make any participant a low-carbon electricity consumer.

\section{Discussion and concluding remarks}
\label{concl}
\paragraph{Experimental results}
We designed and assembled two physical models of indoor farming, greenhouse and vertical farm, to identify and analyze the effects of DR participation on leafy plant growth and the potential economic benefits for the power grid and the industry. The assessment of the actual possibility and conditions of participation in DR for the indoor farming industry necessitates a multidisciplinary approach including IoT, power markets, microelectronics, thermodynamics, and agronomy. To ensure that our work is reproducible, many iterations on the design and corrections on the assembling stage, described in the main text and appendices, were necessary. 

Experiments EI and EII show that participation in either implicit or explicit DR does not adversely affect the vegetative growth stage of the experimental subjects. The behaviour of all varieties used for experiment EIII, shows that the DR condition did not have adverse effects on any of them and we suggest that participation in DR might not have adverse effects on the production of other leafy plants.

\paragraph{Economic impact}
Using our experimental data, we provided an estimation of the economic impact of the industry participation in DR. Though we made several assumptions because of a lack of accurate information from the industry, we are confident that the order of magnitude of the figures we obtained should be correct. We may thus state here that: 

\begin{itemize}
 \item From the industry side, implicit DR represent savings of 15.34\% compared to the base case; explicit DR, 19.44\%; and mixed implicit-explicit DR strategy, 23.03\%. Our results are consistent with the findings reported in \cite{AVGOUSTAKI2021219, AVGOUSTAKI2020107389, ARABZADEH2023120416}. Adding up the figures for the whole industry in ScI, the savings amount to 0.96, 1.2, and 1.44 million USD per year respectively. For ScII, we obtain 10.7, 13.6, and 16.1 million USD per year respectively. 
 \item Currently the indoor farming industry struggles to generate profit not only in Russia \cite{agroinvestor_2021,IKAR2023} but worldwide \cite{Environment2021Stein,Futurefood2017Kurt,Wheat}. Many cultures are not profitable, thus making the scope of the industry narrow  \cite{Environment2021Stein,Futurefood2017Kurt,Wheat,agroinvestor_2021}. The savings evaluated in this work represent a significant reduction in one of the largest production costs of the industry \cite{Futurefood2017Kurt,Verticalfarm2,Wheat}, with almost no technological migration cost compared to other possible solutions such as co-generation, solar or batteries. These savings alone can significantly increase profitability of the industry and hence the industry growth rate. Together with other innovations in automation, lighting technology, and other areas can add-up to open possibilities for including new cultures, and expanding to remote territories where vegetable production is difficult because of harsh weather conditions such as in the Polar regions and other desert regions. 
  \item  The indoor farming industry is often proposed as a more ecological approach to agriculture, with several socio-economic benefits \cite{vanDelden2021,buildings8020024,Verticalfarm2,pennisi2020optimal}, while the high level of energy consumption of the industry is often seen as a drawback \cite{Verticalfarm2,sorensen2016dynagrow,ma2018energy,pennisi2020optimal,Gavhane2023}. In this work, we showed that the  indoor farming industry can provide services to the energy markets, making its energy consumption desirable for the electricity grid and compatible with renewable energy sources.
 \item As regards the benefits from the grid side, adding to the DR program a load of about 55 MW for ScI or about 620 MW for ScII, which following Ref.~\cite{energynet2019} represents about 1\% for ScI and 11\% for ScII of the Russian DR program, entails an economic benefits in ScI of 5.8, 5.9, and 9.9 million USD per year for the implicit DR, explicit DR and mixed DR strategies respectively. For ScII the numbers reach 65.3, 66.5, and 110.9 million USD per year respectively.
 \item For the grid, the indoor farming industry is a very suitable participant for DR, as it can already represent a significant share of the DR program. The indoor farming industry is growing at a rate of about 9\% annually\cite{IKAR2023} in Russia and 10\% worldwide \cite{mordor}, meaning that in a few years it can become the biggest player in the Russian DR program, and a significant one in many other countries. Further, as the aim of indoor farming  is local consumption, the distribution of greenhouses and vertical farms follows the distribution of population, implying that the load is spread across the country offering to the DR program a better nodal control over other possible DR participants. This should be reflected upon to foster savings in transition infrastructure. 
\end{itemize}

\noindent The economic benefits for both the electrical grid and the indoor farming industry are thus significant enough to be perceived as an incentive for participation in implicit and explicit DR.

\paragraph{Carbon footprint reduction}
The assessment of agriculture's sustainability as practices evolve must account for their use of resources, land, water and air pollution, as well as toxicity due to the use of pesticides. In this regards, the indoor agriculture industry could become a promising ecological solutions in terms of sustainability, if its energy efficiency can be increased and its carbon footprint decreased. However, our work shows that by making demand response a standard practice in the Russian indoor agriculture industry, the overall carbon footprint of food production is not necessarily reduced. We thus suggest that integrating the indoor agriculture industry as a provider of services to the energy system together with changes in decarbonisation policies, as well as progress in energy efficiency, may soon transform a power-hungry industry into a desirable asset both for food production and power grid management.

\paragraph{Further development}
The present work can be pursued by developing more complex models for the experimental studies of greenhouses accounting for the heating cost reduction using demand response. As regards the energy consumption optimization and cost reduction, it would be worthwhile to consider co-generation, solar energy, thermal and electric energy storage systems in agricultural complexes, as well as control modelling for greenhouses and vertical farms, and indoor microclimate and light control for plant growth acceleration in both cold and hot climates. 

Our study has shown that the potential of mutual benefit of participation in DR for both the electrical grid and the indoor farming industry is real. However, more work is needed to better quantify these benefits as insufficient power consumption data and information on actual practices in the indoor farming industry precludes the development of more realistic power consumption models and techno-economic studies. Close cooperation with the industry is, for that purpose, of primordial importance.

\section*{CRediT author statement}

\textbf{Javier Penuela :} Conceptualization, Methodology, Software, Formal analysis, Investigation, Writing - Original Draft, Visualization, Writing - Review \& Editing. \textbf{Cécile Ben:} Methodology, Writing - Review \& Editing, Supervision. \textbf{Laurent Gentzbittel:} Methodology, Software, Visualization, Formal analysis, Data curation, Resources, Supervision. \textbf{Henni Ouerdane:} Conceptualization, Methodology, Resources, Writing - Original Draft,  Writing - Review \& Editing,  Supervision.
\section*{Acknowledgements}
H.O. and J.P. acknowledge partial support by the Skoltech program: Skolkovo Institute of Science and Technology – Hamad Bin Khalifa University Joint Projects. J.P. thanks Darya Kostirko for the assistance provided during experiments.

\newpage
 \bibliographystyle{elsarticle-num} 
 \bibliography{refs}
 
 \newpage
 \appendix
\setcounter{table}{0}
\renewcommand\thetable{\Alph{section}.\arabic{table}}
 \section{Experimental setup}
\label{AppA}
The experiment EIII emulates the characteristics of an industrial greenhouse that differ from a vertical farm environment. In EI and EII, each level of the stacked trays is evenly illuminated by LEDs quite close to the plants ($\sim$ 40 cm) while in greenhouses complementary lighting is provided by high-pressure lamps attached to the ceiling to a single layer of crops. To simulate an industrial greenhouse environment we used a phytotron, and for the vertical farm environment we constructed a vertical farm inside the phytotron.
\subsection{Vertical farm} 
\label{AppA1}
The setup consists of one rack fitted with four 1.5$\times$1.0 m$^2$ shelves made of galvanized steel with a weight support capacity of 500 kg per shelf, each shelf separated from the next one above it by a 0.5 m space. An additional shelf at the top serves as a roof for the structure. While the total ground surface area occupied by the setup is 3 m$^2$ (including the surface for the watering system and a servicing area), the setup offers a total of 6-m$^2$ area available for plant growth with controllable lighting, ventilation, temperature, relative humidity and watering systems. We fixed external light-shading screens (made of multiple layers of polyethylene sheets) on the structure to leave enough space for the drainage and ventilation systems, as well as the free movement of one person, while mitigating risks of penetration of parasitic light from the surroundings into the setup. The shelves were waterproofed (with 70 g$\cdot$m$^{-2}$ polyethylene sheets and a food-grade agriculture tray) and thermally insulated to avoid heat transfer effects resulting from the lights below them. The thermal insulation is made of two layers: one 3-cm-thick high-density expanded polystyrene foam with a thermal conductivity of 0.048 W$\cdot$ m$^{-1}\cdot^\circ$C$^{-1}$, and a 3-mm-thin expanded polystyrene reflective (with aluminium foil) thermal insulation sheet with a thermal conductivity of 0.031 W$\cdot$m$^{-1}\cdot ^\circ$C$^{-1}$. Both materials also ensure fire safety. In our lighting fixture, we use SMD 2835 LEDs with a 3:1 red and blue ratio, which is claimed to be optimal for most leafy plants in several works including \cite{pennisi2020optimal,azad2020evaluation}. The LED strips are evenly distributed in the bottom part of each shelf. The LED strips at nominal power can provide a photosynthetic photon flux density (PPFD) of 145$\;\mu $mol$\cdot $sec$^{-1}\cdot $m$^{-2}$ at a  distance of 40 cm. The setup is also equipped with sensing systems for air temperature, air humidity, CO$_2$ concentration. The rack containing the experimental setup was placed inside one phytotron where the ambient humidity, air temperature,  watering pump, and CO$_2$ concentration are monitored and controlled.  Importantly, note that while all the setups' elements are fire-resistant, they accumulate much static charges. In our case, this quickly resulted in the damage of some of the MOSFET-based devices used to control the lights as they are very sensitive to static charges. After replacement, each shelf was grounded, and all the circuits and lights were coated with a thin layer of polyurethane to avoid water damage. 

The water supply had to be organized from top to bottom to reduce the water flow resistance created by the many connectors in the irrigation system. The watering and drainage system are made of several layers, from button up we start with  an agricultural tray with  drainage canals. To reduce the depth of them and therefore accelerate the water drainage a polyethylene sheet covers the surface of the trays. The next layer is composed of capillary mattress capable to absorb up to 3 l$\cdot$ m$^{-2}$ of water; a protective polyethylene sheet with holes is placed on top of them for fungi and algae growth inhibition, finally, the top layer is a drip irrigation system  that distributes water equally. The excess water is eliminated by the drainage system: the agricultural tray collects the water and a hose at the end of the tray guides the water to the drainage of the phytotron.  

The lighting is provided using SMD 2835 LEDs produced by Shenzhen Visva Optoelectronics Co., Ltd. in 2017. We use red and blue colors, in a photon count proportion of 3:1; this ratio was described as optimal for most leafy plants in several articles including \cite{pennisi2020optimal,azad2020evaluation} The LED strips are evenly distributed in the button of the decks, so all the plants receive evenly distributed PPFD. At a height of 40 cm the LED strips at nominal power can provide a PPFD of 145$\;\mu $mol$\cdot $sec$^{-1}\cdot $m$^{-2}$. Each floor uses one 350 W power source with a 12 V DC voltage, the SMD 2835 LED strip lights consume a total of 300 W per floor, and 1.2 kW for the whole setup. Each color of light can be controlled and dimmed independently using pulse-width modulation (PWM) signal from an ESP32 micro-controller on each floor. The goal is to gain flexibility and modularity for future experiments. The PWM signal can be controlled from a local server created to collect the data from all sensors, cameras and control the lights. To reduce losses due to light dispersion, the sides of the setup are covered with a "space blanket" that reflects about 92\% of visible light\cite{article_aluminium}. The PPFD was measured using a DK-PHAR 2.010BS3000TDAC24 quantum sensor from the company DEKA.

Two Waspmotes "plug and sense" sensing boards were placed in different locations to monitor differences in ambient parameters. In case of inconsistencies in the experimental data, the ambient measurements can be used for covariance analysis.The Waspmotes have an integrated battery and measure relative humidity, ambient temperature, CO$_2$ concentration, and air pressure. The measurements are done every five minutes and sent using the XBee communication protocol to a Libelium Meshlium gateway where the data is stored and accessible for the local network.

\subsection{Phytotron} \label{AppA2}
The phytotron is composed of three chambers, the first one with the climate  control system, and the other two are identical growth chambers. with independent microclimate control for each one. The phytotron can provide a wide range of humidity and temperature set points, without CO$_2$  and pressure regulation. The lighting is composed of high pressure sodium lamps and LEDs  (white, red, blue, and far red), the spectrum is controllable as each color of LEDs and the high pressure sodium lamps are independently controlled.  The lights used were DHlicht LED-KE300 with an evenly distributed PPFD of 220$\;\mu $mol$\cdot $sec$^{-1}\cdot $m$^{-2}$ at the soil surface. The plants grow in tables of 250 cm by 140 cm where the watering and drainage is organised analogically to the vertical farm.

\newpage
\setcounter{table}{0}
\setcounter{figure}{0}
\renewcommand\thetable{\Alph{section}.\arabic{table}}
\renewcommand\thefigure{\Alph{section}.\arabic{figure}}    
 \section{Collected data}

\begin{table}[h]
\centering
\begin{tabular}{|l|l|l|l|l|l|}
\hline
\textbf{Species} &
  \textbf{Variety} &
  \textbf{Brand} &
  \textbf{\begin{tabular}[c]{@{}l@{}}Type of \\ seeds\end{tabular}}  &
  \textbf{Season} &
  \textbf{\begin{tabular}[c]{@{}l@{}}Others \end{tabular}} \\ \hline
\begin{tabular}[c]{@{}l@{}}{\it Cichorium} \\ {\it endivia}\end{tabular} &
  \begin{tabular}[c]{@{}l@{}}Blonde \\ à  Coeur\\ Plein\end{tabular} &
  \begin{tabular}[c]{@{}l@{}}La semeuse\\ France\end{tabular} &
  no treatment &
 Summer &
   \\ \hline
\begin{tabular}[c]{@{}l@{}}{\it Lactuca} \\ {\it sativa}\end{tabular} &
  \begin{tabular}[c]{@{}l@{}}Merveille \\ d'Hiver\end{tabular} &
  \begin{tabular}[c]{@{}l@{}}La semeuse\\ France\end{tabular} &
  no treatment  &
Fall   &
  Cold resistant \\ \hline
\begin{tabular}[c]{@{}l@{}}{\it Lactuca} \\ {\it sativa}\end{tabular} &
  Cook RZ &
  \begin{tabular}[c]{@{}l@{}}Rijk  Zwaan\\ Chile\end{tabular} &
  \begin{tabular}[c]{@{}l@{}}Encrustment\\  fungicide \\ ProSeed thiram \end{tabular} &
  Summer &
  \begin{tabular}[c]{@{}l@{}}High\\ temperature \\ resistant\end{tabular} \\ \hline
\begin{tabular}[c]{@{}l@{}}{\it Lactuca} \\ {\it sativa}\end{tabular} &
  Olmetie RZ &
  \begin{tabular}[c]{@{}l@{}}Rijk  Zwaan\\ Chile\end{tabular} &
  \begin{tabular}[c]{@{}l@{}}Encrustment\\  fungicide \\ ProSeed thiram \end{tabular} &
  Summer &
  \begin{tabular}[c]{@{}l@{}}High\\ temperature \\ resistant\end{tabular} \\ \hline
\end{tabular}
\caption{Characteristics of the selected leafy plants.}
\label{tab:plants}
\end{table}

\begin{figure}[h]
    \centering
    \includegraphics[width=\linewidth]{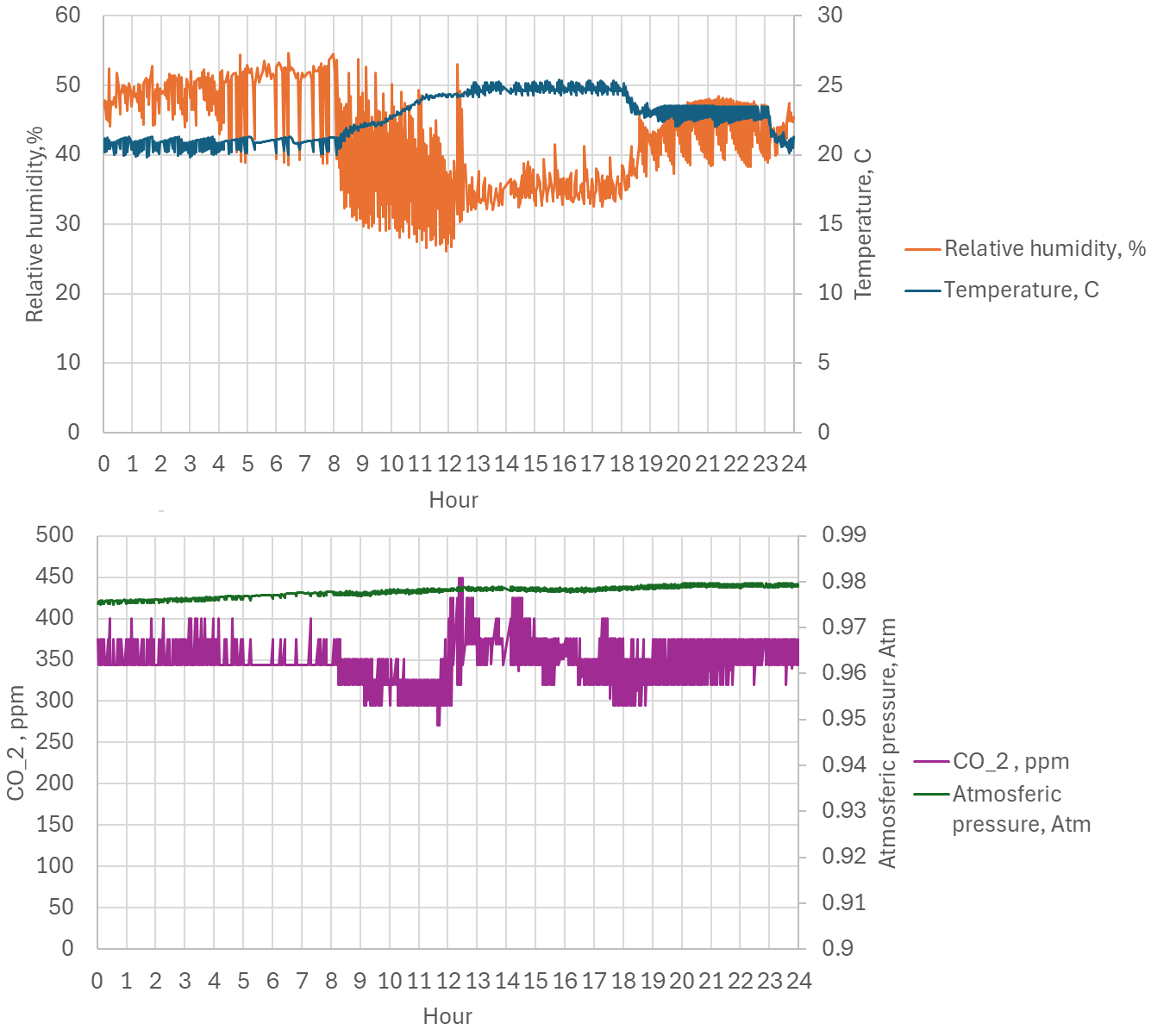}
    \caption{Example of environmental data collected with a resolution of one measurement each four minutes.}
    \label{fig:Sample_data}
\end{figure}

\begin{table}[h]
\begin{tabular}{|l|l|l|l|l|l|}
\hline
\textbf{Variety} &
  \textbf{Condition} &
  \textbf{Plant ID} &
  \textbf{\begin{tabular}[c]{@{}l@{}}Fresh\\ weight (g)\end{tabular}} &
  \textbf{\begin{tabular}[c]{@{}l@{}}Dry\\ weight (g)\end{tabular}} &
  \textbf{\begin{tabular}[c]{@{}l@{}}Quantity\\ of true\\  leaves\end{tabular}} \\ \hline \hline
LS\_cook    & standard & LS\_cook\_std\_44     &  0.31 & 0.03 & 5 \\ \hline
LS\_cook    & standard & LS\_cook\_std\_7      &  0.25 & 0.02 & 5 \\ \hline
LS\_cook    & standard & LS\_cook\_std\_34     &  0.34 & 0.02 & 5 \\ \hline
LS\_cook    & standard & LS\_cook\_std\_23     &  0.27 & 0.02 & 5 \\ \hline
LS\_cook    & standard & LS\_cook\_std\_25     &  0.29 & 0.02 & 5 \\ \hline
Ls\_olmetie & standard & Ls\_olmetie\_std\_123 &  0.33 & 0.01 & 4 \\ \hline
Ls\_olmetie & standard & Ls\_olmetie\_std\_127 & 0.33 & 0.02 & 4 \\ \hline
Ls\_olmetie & standard & Ls\_olmetie\_std\_107 &  0.41 & 0.02 & 4 \\ \hline
Ls\_olmetie & standard & Ls\_olmetie\_std\_138 &  0.31 & 0.02 & 4 \\ \hline
Ls\_olmetie & standard & Ls\_olmetie\_std\_148 & 0.32 & 0.03 & 4 \\ \hline
Ce\_blonde  & standard & Ce\_blonde\_std\_305  & 0.23 & 0.01 & 3 \\ \hline
Ce\_blonde  & standard & Ce\_blonde\_std\_343  & 0.35 & 0.03 & 4 \\ \hline
\end{tabular}
\caption{Example of collected data (all presented measurements were collected on the same day).}
\label{tab:PlantData}
\end{table}

%
%
\begin{table}[h]
\centering
\begin{tabular}{|l|l|l|l|l|l|}
\hline
\textbf{plant ID} &
  \textbf{\begin{tabular}[c]{@{}l@{}}DLI \\ mean\end{tabular}} &
  \textbf{\begin{tabular}[c]{@{}l@{}}Temp. \\ mean\end{tabular}} &
  \textbf{\begin{tabular}[c]{@{}l@{}}CO$_2$\\mean \end{tabular}} &
  \textbf{\begin{tabular}[c]{@{}l@{}}Mean relative\\ humidity\end{tabular}} &
  \textbf{\begin{tabular}[c]{@{}l@{}}mean\\ pressure\end{tabular}} \\ \hline \hline
LS\_cook\_std\_5 & 8.6 mol/m$^2$  & 21.97$^\circ$C & 397.46 ppm& 50.25 \% & 98853 Pa\\ \hline
\end{tabular}
\caption{Example of collected and pre-processed environmental data by the end of the experiment.}
\label{tab:PlantenvData}
\end{table}
\newpage
\setcounter{table}{0}
\renewcommand\thetable{\Alph{section}.\arabic{table}}
\section{Additional results and data}
\label{app:Data}

\begin{table}[]
\footnotesize
\begin{tabular}{|llllllllll|}
\hline
\multicolumn{10}{|c|}{EI and EII}
\\ \hline
\multicolumn{1}{|l|}{} & \multicolumn{3}{l|}{Fresh weight} & \multicolumn{3}{l|}{Dry weight}& \multicolumn{3}{l|}{Number of leaves}\\ 
\hline
\multicolumn{1}{|l|}{Group}& \multicolumn{1}{l|}{Df}  & \multicolumn{1}{l|}{\begin{tabular}[c]{@{}l@{}}Mean\\ Sq\end{tabular}} & \multicolumn{1}{l|}{\begin{tabular}[c]{@{}l@{}}Pr\\ (\textgreater{}F)\end{tabular}} & \multicolumn{1}{l|}{Df}  & \multicolumn{1}{l|}{\begin{tabular}[c]{@{}l@{}}Mean\\ Sq\end{tabular}} & \multicolumn{1}{l|}{\begin{tabular}[c]{@{}l@{}}Pr\\ (\textgreater{}F)\end{tabular}} & \multicolumn{1}{l|}{Df}  & \multicolumn{1}{l|}{\begin{tabular}[c]{@{}l@{}}Mean\\ Sq\end{tabular}} & \begin{tabular}[c]{@{}l@{}}Pr\\ (\textgreater{}F)\end{tabular} \\ \hline
\multicolumn{1}{|l|}{\begin{tabular}[c]{@{}l@{}}Days since sowing \\~\end{tabular}}                   & \multicolumn{1}{l|}{1}   & \multicolumn{1}{l|}{41.67}                                             & \multicolumn{1}{l|}{2.2$\times 10^{-16}$}                                                   & \multicolumn{1}{l|}{1}   & \multicolumn{1}{l|}{37.41}                                             & \multicolumn{1}{l|}{2$\times 10^{-16}$}                                                     & \multicolumn{1}{l|}{1}   & \multicolumn{1}{l|}{28.90}                                             & \textbf{2.2$\times 10^{-16}$}                                                   \\ \hline
\multicolumn{1}{|l|}{Experiment}                                                                     & \multicolumn{1}{l|}{1}   & \multicolumn{1}{l|}{0.479}                                             & \multicolumn{1}{l|}{1.53$\times 10^{-10}$}                                                  & \multicolumn{1}{l|}{1}   & \multicolumn{1}{l|}{0.980}                                             & \multicolumn{1}{l|}{2.2$\times 10^{-16}$}                                                     & \multicolumn{1}{l|}{1}   & \multicolumn{1}{l|}{0.012}                                             & 0.3727                                                         \\ \hline
\multicolumn{1}{|l|}{Condition}                                                                      & \multicolumn{1}{l|}{1}   & \multicolumn{1}{l|}{0.003}                                             & \multicolumn{1}{l|}{0.5971}                                                         & \multicolumn{1}{l|}{1}   & \multicolumn{1}{l|}{0.011}                                             & \multicolumn{1}{l|}{0.334}                                                          & \multicolumn{1}{l|}{1}   & \multicolumn{1}{l|}{0.022}                                             & 0.2137                                                         \\ \hline
\multicolumn{1}{|l|}{\begin{tabular}[c]{@{}l@{}}Days since\\ sowing:Experiment\end{tabular}}         & \multicolumn{1}{l|}{1}   & \multicolumn{1}{l|}{0.738}                                             & \multicolumn{1}{l|}{1.69$\times 10^{-14}$}                                                  & \multicolumn{1}{l|}{1}   & \multicolumn{1}{l|}{1.044}                                             & \multicolumn{1}{l|}{2.2$\times 10^{-16}$}                                                     & \multicolumn{1}{l|}{1}   & \multicolumn{1}{l|}{0.594}                                             & 2.$\times 10^{-9}$                                                    \\ \hline
\multicolumn{1}{|l|}{Residuals}                                                                      & \multicolumn{1}{l|}{135} & \multicolumn{1}{l|}{0.010}                                             & \multicolumn{1}{l|}{}                                                               & \multicolumn{1}{l|}{135} & \multicolumn{1}{l|}{0.011}                                             & \multicolumn{1}{l|}{}                                                               & \multicolumn{1}{l|}{135} & \multicolumn{1}{l|}{0.014}                                             &                                                                \\ \hline
\multicolumn{10}{|c|}{EIII}                                                                                                                                                                                                                          \\ \hline
\multicolumn{1}{|l|}{\begin{tabular}[c]{@{}l@{}}Days since \\ sowing\end{tabular}}                   & \multicolumn{1}{l|}{1}   & \multicolumn{1}{l|}{52.05}                                             & \multicolumn{1}{l|}{2.2$\times 10^{-16}$}                                                   & \multicolumn{1}{l|}{1}   & \multicolumn{1}{l|}{51.88}                                             & \multicolumn{1}{l|}{2.2$\times 10^{-16}$}                                                   & \multicolumn{1}{l|}{1}   & \multicolumn{1}{l|}{43.11}                                             & 2.2$\times 10^{-16}$                                                  \\ \hline
\multicolumn{1}{|l|}{Variety}                                                                        & \multicolumn{1}{l|}{3}   & \multicolumn{1}{l|}{0.057}                                             & \multicolumn{1}{l|}{0.0067}                                                         & \multicolumn{1}{l|}{3}   & \multicolumn{1}{l|}{0.074}                                             & \multicolumn{1}{l|}{0.0041}                                                         & \multicolumn{1}{l|}{3}   & \multicolumn{1}{l|}{3.957}                                             & 2.2$\times 10^{-16}$                                                   \\ \hline
\multicolumn{1}{|l|}{Condition}                                                                      & \multicolumn{1}{l|}{1}   & \multicolumn{1}{l|}{0.077}                                             & \multicolumn{1}{l|}{0.0181}                                                         & \multicolumn{1}{l|}{1}   & \multicolumn{1}{l|}{0.174}                                             & \multicolumn{1}{l|}{0.0012}                                                         & \multicolumn{1}{l|}{1}   & \multicolumn{1}{l|}{0.190}                                             & 0.00011                                                      \\ \hline
\multicolumn{1}{|l|}{\begin{tabular}[c]{@{}l@{}}Days since\\ sowing:Variety\end{tabular}}            & \multicolumn{1}{l|}{3}   & \multicolumn{1}{l|}{0.048}                                             & \multicolumn{1}{l|}{0.0147}                                                         & \multicolumn{1}{l|}{3}   & \multicolumn{1}{l|}{0.063}                                             & \multicolumn{1}{l|}{0.0099}                                                         & \multicolumn{1}{l|}{3}   & \multicolumn{1}{l|}{0.324}                                             & 6$\times 10^{-14}$                                                   \\ \hline
\multicolumn{1}{|l|}{\begin{tabular}[c]{@{}l@{}}Days since\\ sowing:Condition\end{tabular}}          & \multicolumn{1}{l|}{1}   & \multicolumn{1}{l|}{0.004}                                             & \multicolumn{1}{l|}{0.5702}                                                         & \multicolumn{1}{l|}{1}   & \multicolumn{1}{l|}{0.009}                                             & \multicolumn{1}{l|}{0.4509}                                                         & \multicolumn{1}{l|}{1}   & \multicolumn{1}{l|}{0.010}                                             & 0.35984                                                        \\ \hline
\multicolumn{1}{|l|}{Variety:Condition}                                                              & \multicolumn{1}{l|}{3}   & \multicolumn{1}{l|}{0.010}                                             & \multicolumn{1}{l|}{0.5271}                                                         & \multicolumn{1}{l|}{3}   & \multicolumn{1}{l|}{0.008}                                             & \multicolumn{1}{l|}{0.6630}                                                         & \multicolumn{1}{l|}{3}   & \multicolumn{1}{l|}{0.049}                                             & 0.00792                                                        \\ \hline
\multicolumn{1}{|l|}{\begin{tabular}[c]{@{}l@{}}Days since sowing:\\ Variety:Condition\end{tabular}} & \multicolumn{1}{l|}{3}   & \multicolumn{1}{l|}{0.006}                                             & \multicolumn{1}{l|}{0.7135}                                                         & \multicolumn{1}{l|}{3}   & \multicolumn{1}{l|}{0.012}                                             & \multicolumn{1}{l|}{0.5386}                                                         & \multicolumn{1}{l|}{3}   & \multicolumn{1}{l|}{0.006}                                             & 0.68287                                                        \\ \hline
\multicolumn{1}{|l|}{Residuals}                                                                      & \multicolumn{1}{l|}{144} & \multicolumn{1}{l|}{0.013}                                             & \multicolumn{1}{l|}{}                                                               & \multicolumn{1}{l|}{144} & \multicolumn{1}{l|}{0.016}                                             & \multicolumn{1}{l|}{}                                                               & \multicolumn{1}{l|}{144} & \multicolumn{1}{l|}{0.012}                                             &                                                                \\ \hline
\end{tabular}
\caption{Results of two-way ANOVA analysis for EI, EII and EIII using the statistical models \ref{eq:AnovaEI_EII} and \ref{eq:AnovaEIII} for the measured biological parameters (fresh weight, dry weight and number of true leaves). In the table, values of the degrees of freedom (Df), the mean sum of squares (Mean Sq) and the p-value (Pr($\geq$ F)). The interaction between independent factors is shown as Factor1: Factor2.}
\label{tab:ANOVA}
\end{table}

\begin{table}[h]
\centering
\begin{tabular}{|l|l|l|}
\hline
\textbf{\begin{tabular}[c]{@{}l@{}}Month\end{tabular}} &
  \textbf{\begin{tabular}[c]{@{}l@{}}Monthly base case \\ electricity cost\end{tabular}} &
  \textbf{\begin{tabular}[c]{@{}l@{}}Monthly implicit DR\\ electricity cost\end{tabular}} \\ \hline\hline
01          & 862,293 & 736,037 \\ \hline
02          & 678,282 & 575,015 \\ \hline
03          & 529,891 & 455,300 \\ \hline
04          & 323,073  & 274,521 \\ \hline
05          & 89,764       & 77,340       \\ \hline
06          & 0       & 0       \\ \hline
07          & 15,774       & 13,808       \\ \hline
08          & 175,462  & 110,121  \\ \hline
09          & 477,189 & 405,468 \\ \hline
10          & 688,435 & 576,382 \\ \hline
11          & 865,368 & 739,334 \\ \hline
12          & 856,704 & 745,323 \\ \hline
\end{tabular}
\caption{Monthly energy costs per hectare (rub/month$\cdot$ha) for the base case simulation and implicit DR participation. As of 2024 the USD/RUB conversion rate is assumed to be 1/90.}
\label{tab:savingspermonth}
\end{table}

\end{document}